\documentclass{aa}

\usepackage{graphicx}
\usepackage{txfonts}
\usepackage{longtable}
\usepackage{comment}

\usepackage[colorlinks=true, citecolor=blue, linkcolor=black, urlcolor=blue]{hyperref}

\usepackage[dvipsnames]{xcolor}

\begin{document}

   \title{Exploring Low-Amplitude Variability in First Overtone Cepheids with TESS}
\titlerunning{TESS 1O Cepheids}

   \author{E. Plachy
          \inst{1}\fnmsep\thanks{eplachy@konkoly.hu}, 
         H. Netzel\inst{2}
          \and
         A. B\'odi\inst{3}
          }

   \institute{Konkoly Observatory, HUN-REN Research Centre for Astronomy and Earth Sciences,
              MTA Centre of Excellence, \\ Konkoly Thege u. 15-17., 1121 Budapest, Hungary
            \and
            Nicolaus Copernicus Astronomical Center, Bartycka 18, 00-716 Warsaw, Poland
            \and
            Department of Astrophysical Sciences, Princeton University 
             Peyton Hall, 4 Ivy Lane, Princeton, NJ 08544, USA
             }


 
  \abstract
   {Classical Cepheid stars that pulsate in the first overtone radial mode often exhibit additional periodicities at the millimagnitude level. Extensive studies of the OGLE data of the Magellanic Clouds have revealed distinct groups based on their period ratio with the first overtone mode. These groups are similar to those found in overtone RR Lyrae stars. Theoretical calculations suggest that some of the observed periodicities may be consistent with non-radial modes, while others remain unexplained.  Currently, we only know of a handful of examples from the Galactic Cepheid sample that exhibit low-amplitude periodicities.}
   {
   The purpose of this study is to undertake a systematic search for low-amplitude variability in overtone Cepheids of the Milky Way in the photometric data of the full-frame images of the Transiting Exoplanet Survey Satellite, which were produced with the MIT Quick Look Pipeline.
     
   }
   {We applied standard Fourier analysis and classified the additional signals  according to their period ratio to the overtone pulsation period.}
   {We found 127 stars in total to exhibit additional periodicities. In 17 stars, these can be identified as a second radial overtone. A further 83 stars were observed to display periodic signals with a ratio of $P_{\mathrm{x}}/P_{1\mathrm{O}}$ in the range \mbox{0.60$-$0.65}. In 15 stars, the $P_{1\mathrm{O}}/P_{\mathrm{x}}$ is found to be near $\sim$0.68, of which six are also found to be in the previous group. Furthermore, we observed the presence of low-amplitude signals in 22 stars outside the aforementioned period ratios. It is possible that some of these may be direct detections of non-radial modes, with no harmonic frequency peak in the  \mbox{0.60$-$0.65} period range.  
   
   }
   {Most of the properties of the additional signals detected in this study are consistent with previous findings regarding Cepheids in the Magellanic Clouds. However, TESS measurements revealed that the amplitudes and frequencies of these signals often vary within a TESS sector, a phenomenon that challenges theoretical models. We also found that reliable analysis of these signals with TESS is possible for a moderate fraction of overtone Cepheids. The primary constraints are the limited extent of the data and the relatively poor photometric quality in some Galactic fields. Further careful corrections to the light curves may improve detectability.}


   \keywords{Stars: variables: Cepheids --
                Techniques: photometric
               }

   \maketitle
   
%

\section{Introduction}

Classical Cepheids play an important role in stellar astrophysics. They are primary distance indicators due to their period-luminosity relations and are also key objects in stellar evolution and pulsation. While most classical Cepheids pulsate in a single radial mode, either the fundamental or the first overtone,  multi-mode pulsations up to the third overtone are not uncommon. Over the past decade, discoveries have revealed the existence of low-amplitude additional periodicities in a significant number of classical Cepheids that cannot be explained by radial pulsation. 

The first such signals were reported by \citet{Moskalik2004} in overtone Cepheids of the Large Magellanic Clouds observed by the Optical Gravitational Lensing Experiment (OGLE, \citealt{ogle-1992}). In subsequent years, further discoveries were made, including the detection of additional signals in double-mode Cepheids \citep{Moskalik2008, Moskalik2009}. It was soon realized that these additional frequencies form sequences in the Petersen diagram \citep{petersen} with the period ratio ($P_{\mathrm{X} } / P_{1\mathrm{O}}$) in the range \mbox{0.60$-$0.65}.  
\citep{sos2008,sos2010}. An extensive investigation has revealed another group that has a ratio of $P_{\mathrm{X} }/ P_{1\mathrm{O}}$$\sim$1.46 (more often referred to as its reciprocal: $P_{1\mathrm{O}} / P_{\mathrm{X}} $$\sim$0.68 \citep{suveges}. The first member of the latter group was discovered by \citet{poretti2014} in data from the CoRoT mission \citep{Corot}.

The two groups, designated as the 0.61 and 0.68 groups, were also identified in overtone RR Lyrae (RRc) stars with similar frequency sequences for the former case (see \citealt{Netzel2019} and references therein). Moreover, the 0.61 signal has also been found in an overtone anomalous Cepheid, XZ Cet \citep{plachy2021}, it being the only example of this type so far. In this article, the notation $f_{0.61}$ and $f_{0.68}$ will be used, but it should be emphasized that while $f_{0.61}$ has the higher frequency (shorter period), $f_{0.68}$ has the lower frequency (longer period) in relation to the first overtone mode.

Since the discovery of these additional periodicities, non-radial pulsation has been suspected to be their origin. Similarly, weak non-radial modes were suggested as the cause of phase curve changes observed in some overtone Cepheids \citep{szabados-2009}. Detailed theoretical calculations for the 0.61 signals were first provided by \citet{dziem}. In these non-radial pulsation models, the sequences correspond to the harmonics of the modes with spherical degrees of $\ell=7,8,9$ in classical Cepheids and $\ell=8,9$ in RR Lyrae stars. In RR Lyrae stars, the third sequence consists of combinations of frequencies between these two. The true frequencies in the model are at the 1/2$f_x$ subharmonics, which may not appear with detectable amplitudes due to geometric cancellations. In fact, subharmonics are present in only a fraction of stars with 0.61 modes \citep{Smolec2016}. The theory predicts the $\ell=10$ sequence with a period ratio slightly lower than 0.60. Sequences with this ratio have recently been discovered in RR Lyrae stars by \cite{k2_rrc} and \cite{benko-2023} using space-based photometry from the K2 \citep{k2} and Transiting Exoplanet Survey Satellite (TESS, \citealt{Ricker2015}) missions, respectively. To date, no comprehensive theoretical explanation has been proposed for the 0.68 group. Moreover, the complexity of the issue has increased with the recent discovery that the 0.68 period ratio also presents in fundamental mode RR Lyrae stars \citep{rrab068}. 

In contrast to RR Lyrae stars, where space-based data have been used to search for low-amplitude signals, the majority of Cepheid discoveries have been associated with the OGLE survey. The current number of overtone and double-mode Cepheids with additional frequencies in the Magellanic system is now over two thousand, as recently reported by \citet{smolec2023}. This analysis clearly shows the presence of a third group close to the period of the overtone mode, as originally suggested in the work of \cite{oliwia2020}. Explaining these signals is also challenging, as the low-order non-radial modes expected in this region are damped \citep{Mulet2007}. Nevertheless, they differ from the periodic modulation peaks detected in some overtone Cepheids. 

Our understanding of the occurrence of additional signals in Milky Way Cepheids is more limited. The OGLE mission also provides observations in the Galactic disk and bulge, but much fewer Cepheids with additional signals have been found in these regions \citep{pietr2013,Rathour2021}. The discrepancy is explained with a significantly higher detection limit towards the Galactic fields due to the lower sampling rates and shorter baseline of observations. Spectroscopic detection of the 0.61 and 0.68 signals is available only for bright Galactic Cepheids \citep{veloce1}. The aim of this study is to increase the number of known Galactic Cepheids with additional signals by analyzing the data of the TESS full-frame images (FFI).

\section{Data and method}

As demonstrated by previous studies using data from the TESS mission, continuous space photometry is highly effective in detecting low-amplitude features, including additional pulsation modes in RR Lyrae and Cepheid stars \citep{molnar2022,benko-2023,plachy2021}. However, an extensive search for additional modes in overtone Cepheids using TESS has not yet been carried out. In this section, we describe the characteristics of the TESS data and explain how quality issues can restrict the selection and analysis of targets.

\subsection{The properties of TESS data}
\label{tessdata}

TESS was designed to collect continuous photometry from almost the entire sky in 27-day sectors. Its field of view spans an area of $24 \times 96$ degrees and covers the sky in overlapping sectors that rotate around the ecliptic poles. Each sector corresponds to two orbits of the satellite around Earth. This design has several implications for the data. When the spacecraft reaches the perigee, it pauses its observations in order to downlink the accumulated data. This results in a data gap appearing in the middle of the light curves. Scattered light also introduces systematic trends, which are particularly noticeable during certain orbital phases and can vary depending on the specific sector and camera (for a detailed description of this effect, see the TESS Instrument Handbook\footnote{\url{https://archive.stsci.edu/files/live/sites/mast/files/home/missions-and-data/active-missions/tess/_documents/TESS_Instrument_Handbook_v0.1.pdf}}).  

Although correction methods can significantly reduce the impact of scattered light, the effectiveness can vary depending on the specific characteristics of the target. For high-amplitude pulsators, there is often an average brightness difference between the data from the two orbits. Fortunately, this can be easily corrected by shifting the data. However, if the scattered light was not fully corrected by the photometric pipeline, further detrending is required. Due to the discontinuity, the two orbits of the sectors must be treated separately when detrending. In their study of TESS RRc stars, \citet{benko-2023} used polynomial fitting via phase dispersion minimization (see also \citealt{Bodi2022}). Unlike RRc stars, overtone Cepheids can have longer periods (up to 6 days), resulting in a small number of observed pulsation cycles per orbit. 
Furthermore, additional modes manifest themselves as cycle-to-cycle variations in the light curve. Such variations can be distorted or completely eliminated by a detrending algorithm. 
Therefore, we decided not to use detrending, but only to correct for the average difference in brightness between the orbits.
The measured pulsation amplitudes may also differ between sectors due to changes in position on the detector and the level of contamination. This was also corrected by shifting when data from different sectors were combined.

Even though TESS has a brightness limit of approximately 16 magnitude, the detection of millimagnitude signals becomes uncertain beyond 14--15 mag due to the increased noise level. 
Consequently, we focus on brighter Cepheids. In recent years, several light curve products and reduction pipelines have become available for processing FFI targets. For this research, we decided to use the light curves from the MIT Quick Look Pipeline (QLP, \citealt{qlp-0,qlp-1,qlp-2}). This processes FFI targets up to magnitude 13.5, and of all the pipelines, this one has processed the most sectors at the time of writing this paper. Light curves are available in the MAST Archive\footnote{\url{https://archive.stsci.edu}}. 

We used photometric QLP data up to Sector 81 (S81). In terms of sampling, this set of data is not uniform. FFIs were taken with a 30-minute cadence in the two years of the Primary Mission (S1--S26), then reduced to 10 minutes in the First Extended Mission (S27--S55) and further reduced to 200 seconds in the Second Extended Mission (from S56). Because a large part of the sky was repeatedly observed in the extended missions, many of the targets were observed at least three times. Several specific regions of the Galactic plane were observed in consecutive sectors. This provides an opportunity to combine successive data sets and improve frequency resolution. 

\subsection{Target selection and data preparation}
To select targets for our analysis we rely on the
catalog\footnote{\url{https://www.astrouw.edu.pl/ogle/ogle4/OCVS/allGalCep.listID}} of \citet{pietr}, the most complete and regularly updated list of verified classical Cepheids in the Milky Way collected from various sky surveys and the literature. The version updated in March 2025 is used here. The list contains 3644 classical Cepheids of which 1078 pulsate in the first overtone radial mode. We found QLP data in Sectors 1--81 for 495 stars from the overtone Cepheid group. We used the normalized Single Aperture Flux (SAP\_FLUX) data for analysis. Upon inspection of the light curves, it became clear that only a limited sample met our data quality requirements. Noise, trends, contamination, and an insufficient number of observed cycles were the main problems. The latter was due to two reasons, either because the period of the star was too long or because too many poor-quality segments had to be cropped. We found that the spline-detrended data provided by the QLP pipeline (called KSPSAP\_FLUX or DET\_FLUX in later sectors) do not solve our trend problems. To filter out poor-quality measurements, only the data points with the 0 quality flag were used. Stars with periods longer than 5 days were only examined if data from consecutive sectors could be merged. In data stitching, we used the average value of the peak-to-peak amplitudes as a reference value.

Due to the data quality problems mentioned above, a large fraction of the sample stars were excluded, leaving 301 stars for analysis.

\subsection{Analysis}

\begin{figure}
\centering
\includegraphics[width=0.49\textwidth]{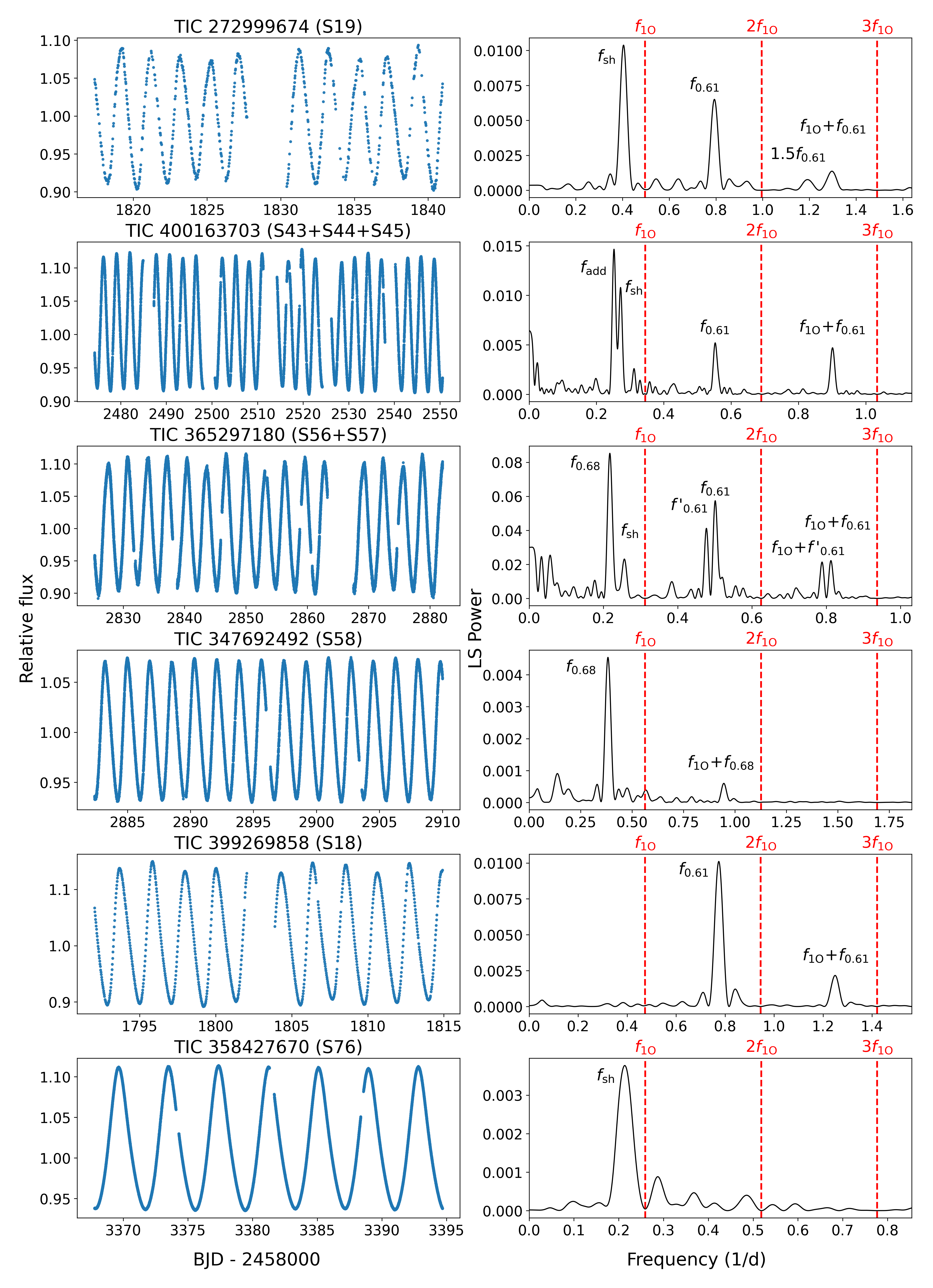}
\caption{Example TESS QLP light curves and residual spectra for Cepheids with additional signals. Note that the frequency ranges on the right panels are different. 
Red dashed lines mark the positions of the subtracted frequencies of the first overtone mode ($f_{\rm{1O}}$, 2$f_{\rm{1O}}$, 3$f_{\rm{1O}}$). The low-amplitude peaks identified are labelled. $f_{\rm{sh}}$ denotes the subharmonic frequency of the $f_{0.61}$ peak. 
}
\label{fig:examples}
\end{figure}

We perfomed Standard Fourier analysis using the open source \texttt{python} code developed by \cite{bodi2024}. To reveal the low-amplitude additional frequencies, we prepared residual spectra by prewhitening with the main frequency and its harmonics. Our frequency range of interest extends up to the second harmonic. 
The low-amplitude frequency peaks found in the residual spectra were grouped according to their ratio with the overtone mode. We also searched for combination, harmonic, and subharmonic frequency peaks. We accepted additional frequencies as significant at a signal-to-noise ratio greater than 4.0. The Fourier amplitudes of these peaks, calculated from the relative flux, are usually greater than 0.001, except for the highest-quality light curves where peaks below this limit can be identified.  The errors were estimated analytically using the formulae of \citet{errors}. We also used the \texttt{Period04} tool \citep{period04} to verify ambiguous frequency peaks near the significance level.

We analyzed the data from each sector individually, as well as the combined data of successive sectors. Due to the large gap between the observations, we did not merge the data from different years together. We also did not merge data from different sectors if there was a significant difference in pulsation amplitudes or data quality. 

In Fig.~\ref{fig:examples} we show example TESS QLP light curves and residual spectra with additional periodicites.

\section{Results}

In this section we present our results on the different types of low-amplitude periodicities. We present a set of stars in which we have found the second radial overtone (2O). We report new members for the 0.61 and 0.68 groups. We also discuss the additional frequencies found outside the ranges of the above groups, including the possible direct detection of non-radial modes without the harmonic frequency near the 0.61 ratio. Table \ref{tab:sum} summarizes our findings. Table \ref{tab:targets} includes the full list of stars with additional periodicities. Tables with the pulsation properties can be found in the Appendix.

\begin{table}[h!]

    \centering
    \begin{tabular}{ll}
Type & Number \\  
\noalign{\vskip 2pt} 
\hline
\hline
\noalign{\vskip 2pt} 
1O2O & 17 \\
\noalign{\vskip 2pt} 
\hline
\noalign{\vskip 2pt} 
0.61 stars in total &  83 \\
0.61 stars with subharmonics & 78 \\
\noalign{\vskip 2pt} 
\hline
\noalign{\vskip 2pt} 
0.68 stars in total & 15 \\
simultaneous 0.68 and 0.61 frequencies  & 6 \\
\noalign{\vskip 2pt} 
\hline
\noalign{\vskip 2pt} 
stars with other periodicites in total & 22 \\
candidate direct detection of non-radial signals  & 11 
\\
\noalign{\vskip 2pt}

\noalign{\vskip 2pt} 
\hline
\noalign{\vskip 2pt} 

    \end{tabular}
    \caption{ Summary table showing the number of stars found in different groups of additional periodicities.}
    \label{tab:sum}
\end{table}

\begin{table*}[ht!]
    \centering
    \begin{tabular}{ccccccc}
    TIC ID &  Gaia DR3 ID & Other ID & RA [deg] & Dec [deg] & $T_{\rm{mag}}$ & Group    \\
    \hline

4712990	&	2930138465868534528	&	ASASSN-V  J071753.64-195040.2	&	109.473505	&	-19.844523	&	12.827 	&	0.61	\\
4867143	&	5618219442576595968	&	OGLE-GD-CEP-1326	&	109.790976	&	-22.342993	&	12.3172 	&	0.61	\\
27344118	&	232773865804463232	&	NSVS 4261418	&	65.648572	&	45.581916	&	11.1202	&	0.61	\\
28184016	&	253367890394497408	&	ASASSN-V J042921.05+435914.6	&	67.337709	&	43.987392	&	13.3363	&	0.61	\\
28248697	&	253825115433087360	&	ASASSN-V J042922.58+453404.2	&	67.344249	&	45.567552	&	11.4301	&	0.68	\\

\vdots	&	\vdots	&	\vdots	&	\vdots	&	\vdots &	\vdots	& \vdots	\\
\hline
    \end{tabular}
    \caption{Name list of stars with additional periodicities. Coordinates and TESS magnitude are included. The full table listing 127 stars table can be found at the CDS.}
    \label{tab:targets}
\end{table*}

\subsection{Additional radial modes and amplitude modulations}

\begin{figure}
\centering
\includegraphics[width=0.49\textwidth]{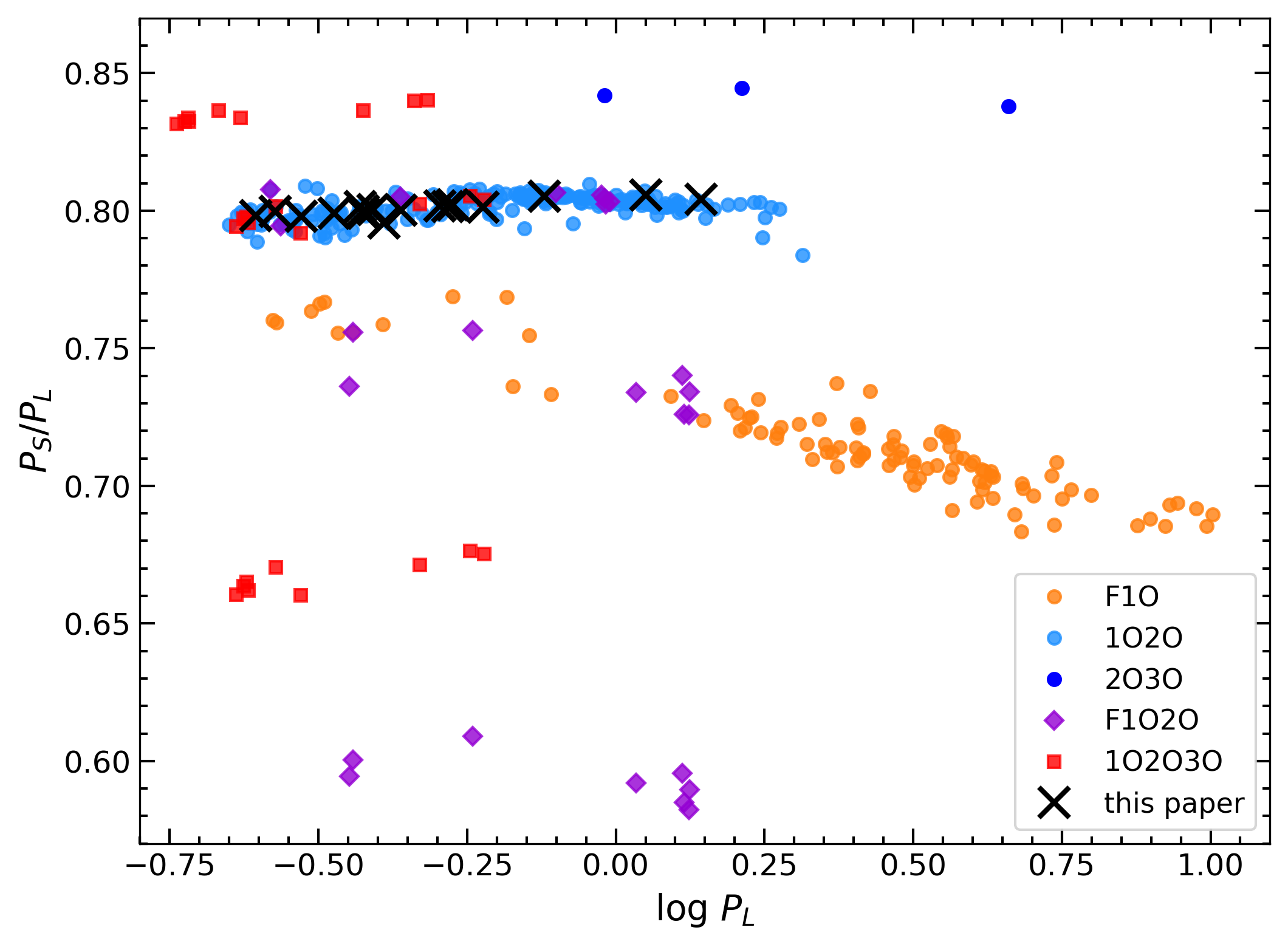}
\caption{Petersen diagram of radial multimode classical Cepheids in the Milky Way. Colored dots are drawn in from the catalog of \citet{pietr} and from the the candidates of \citet{Rathour2021}. The black crosses show the double mode stars found in this study.}
\label{fig:petersen_rad}
\end{figure}

Seventeen stars of our sample show additional variation identified as the second radial mode based on their period ratios and positions on the Petersen diagram (Fig.~\ref{fig:petersen_rad}). Their pulsation properties are listed in Table \ref{tab:dm}. Combination frequencies between 1O and 2O are detected in all cases. 

Most of these stars have been observed in multiple sectors of different years, providing insight into temporal changes in pulsation. Looking at the A$_{\rm{2O}}$/A$_{\rm{1O}}$ amplitude ratios in Table \ref{tab:dm}, we can see that there are significant differences between sectors for most of them, mainly due to the amplitude variation of the 2O.

\begin{figure*}
\centering
\includegraphics[width=\textwidth]{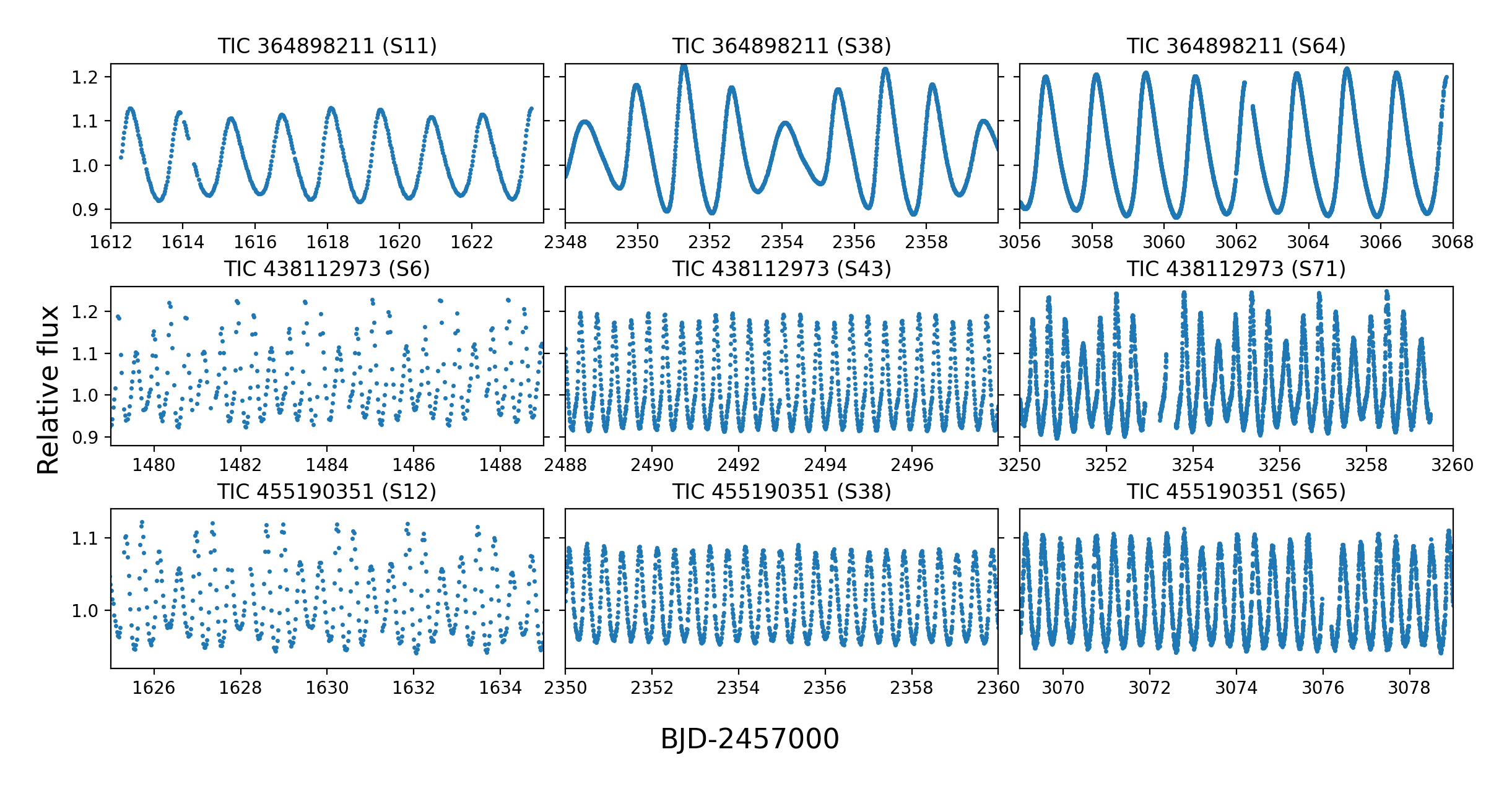}
\caption{Light curves segments are presented for three 1O2O stars that exhibit visible change in the amplitude ratios. The three columns represent the primary and the two extended missions. The sectors from which the data is adopted are indicated in the brackets.}
\label{fig:ampvar}
\end{figure*}

There are three stars in the sample where this change is extremely large (Fig.~\ref{fig:ampvar}). The amplitude ratios in TIC~364898211 (MS~Mus) and TIC~438112973 (GM~Ori) change by a factor of more than ten. The existence of the 2O overtone has been reported for both stars (see \citealt{Khruslov2013} and \citealt{Rathour2021}), but was later not detected in the Gaia DR3 data \citep{ripepi2022}. The third star in Fig \ref{fig:ampvar}, TIC~455190351 (OGLE~GD-CEP-1869), shows the largest amplitude variation (A$_{\rm{2O}}$/A$_{\rm{1O}}$ changes by a factor of $\sim$30). We reanalyzed the OGLE \textit{I}--band photometry for this star and found that it was not possible to recover the 2O due to the daily aliases.

The three stars presented above are likely members of the 1O2O group exhibiting high-amplitude periodic modulation. This phenomenon was reported by \citet{Moskalik2004} and \citet{Moskalik2009} and more recently by \citet{smolec2023} in Cepheids of the Magellanic Clouds. The group is characterized by anti-correlated amplitude and phase modulations with periods over 700 days. The variations detected in these stars are similar to the mysterious phenomenon commonly seen in RR Lyrae stars, called Blazhko effect \citep{Blazhko}. With data from the future TESS cycles, we hope to learn more about these stars and perhaps even determine the modulation periods.

Another important question besides modulation is whether double radial mode Cepheids can exhibit similar additional periodicities to their 1O counterparts. Such stars have already been discovered in the Magellanic System \citep{smolec2023}: nineteen F1O stars with the $f_{0.61}$ frequency and four with $f_{0.68}$, of which one shows both types of additional variability. One example of a 1O2O and another of a 1O3O Cepheid with additional $f_{0.61}$ frequency are also known. We were unable to find any signals in our double-mode stars that could be attributed to $f_{0.61}$ or $f_{0.68}$. The majority of the low-amplitude frequency peaks can be explained by a linear combination of the two radial modes. We found other weak signals appearing at seemingly random frequencies in roughly half of the sample. We suspect that these are due to contamination. We also detected residual power around the radial modes after prewithening, which is likely due to the amplitude variations in the stars discussed above.

\subsection{The 0.68 group}

We found fifteen stars where a clear additional signal at $P_{1\mathrm{O}} / P_{\mathrm{X}} $$\sim$0.68 is visible (see Table \ref{tab:068}). Of these, fourteen are new discoveries. The $f_{0.68}$ signal was detected in TIC~197776441 (V0411 Lac) by \citet{veloce1} in spectroscopic time series. Our analysis confirms the visibility of this periodicity in the photometric data of the star. 

Six stars from the 0.68 group are also members of the 0.61 group. We did not find the combination frequency between these two types of signals in either cases. The combination frequency with the first overtone mode ($f_{1\mathrm{O}}$+$f_{0.68}$) appears in only two stars, in TIC~307298395 and TIC~239378995. This latter star shows other additional periodicities too (see section \ref{sec:other}).  The remaining seven stars of the sample show no additional periodicities apart from $f_{0.68}$.

\begin{figure}
\centering
\includegraphics[width=0.48\textwidth]{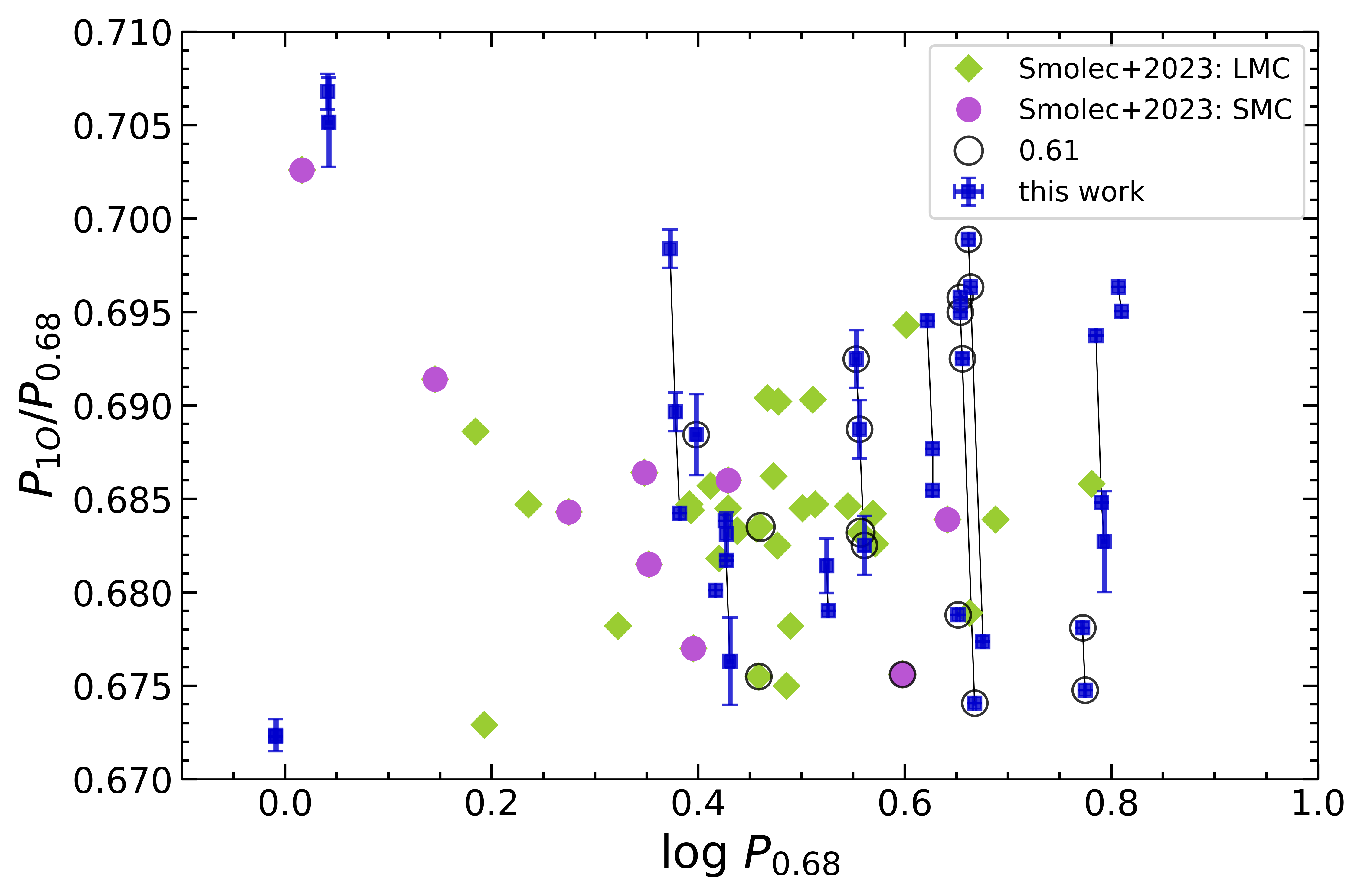}
\caption{Petersen diagram of the 0.68 stars found in this work (blue squares) plotted onto the results of \citet{smolec2023} found in the same range of period ratio (green diamonds for LMC and purple dots for SMC).  The values derived for the same stars, but from different sectors are connected with black lines. }
\label{fig:68}
\end{figure}

We calculated the period ratio for the fifteen stars in different sectors and plotted the results in Fig.~\ref{fig:68}, along with the period ratios found in the 0.67--0.70 range for stars in the Magellanic Clouds, determined by \citet{smolec2023}, for comparison. 
Both sets of stars show similar scatter in the $P_{1\mathrm{O}} / P_{0.68}$ values, but for the stars in the TESS sample, the period ratios derived from the different sectors differ significantly for some stars. The connected symbols refer to the same star in Fig.~\ref{fig:68}. 
It is not clear whether this is an intrinsic effect or due to the possible underestimation of the errors. However, if the $f_{0.68}$ frequency varies over time, this is inconsistent with the findings of \citet{smolec2023}, who reported $f_{0.68}$ to be coherent. We may speculate that difference in metallicity between the Cepheids of the Magellanic System and the Milky Way may be a contributing factor.

\subsection{The 0.61 group}
\label{sec:061}

The search for additional frequencies at P$_{\mathrm{X}}$/$P_{1\mathrm{O}}$$\sim$0.6--0.65 resulted in 83 stars. It is quite common that these form a combination frequency with the overtone mode, the presence of which confirms the detection (marked with 'c' in Table \ref{tab:061}). Among the combinations, $f_{\rm{1O}}$+$f_{0.61}$ is always present, with $2f_{\rm{1O}}$+$f_{0.61}$  and $f_{0.61}$--$f_{\rm{1O}}$ being occasionally visible (mostly a function of data quality, not specifically marked in Table \ref{tab:061}). The subharmonic frequency ($f_{\rm{sh}}$$\sim$0.5$f_{0.61}$) appears in almost all stars in at least one sector. We found only five stars where a significant peak could not be identified in the subharmonic region, in any sector. In certain instances, only power excess is apparent, rather than a subharmonic peak (this is marked with 'p'). 

\begin{figure}
\centering
\includegraphics[width=0.49\textwidth]{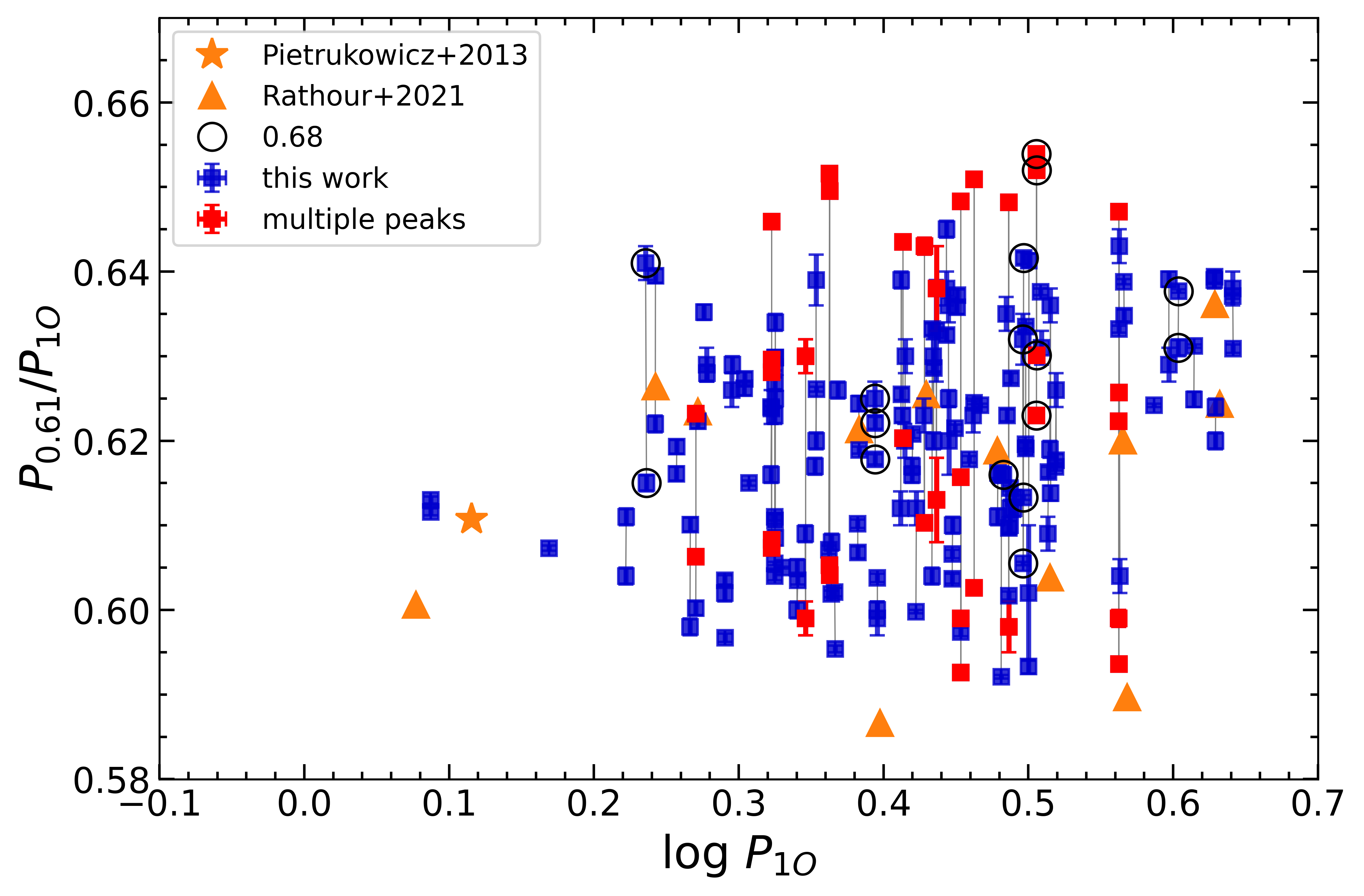}
\caption{Petersen diagram of the 0.61 group. The previously discovered Galactic members of the group \citep{pietr2013,Rathour2021} are marked with orange asterisk and triangles. Black lines connect the values for the same star. 
Red symbols denote stars for which we found multiple peaks for the 0.61 frequency group in the same light curve. The stars which are also member of the 0.68 group are marked with circles.}
\label{fig:petersen}
\end{figure}

\begin{figure}
\centering
\includegraphics[width=0.49\textwidth]{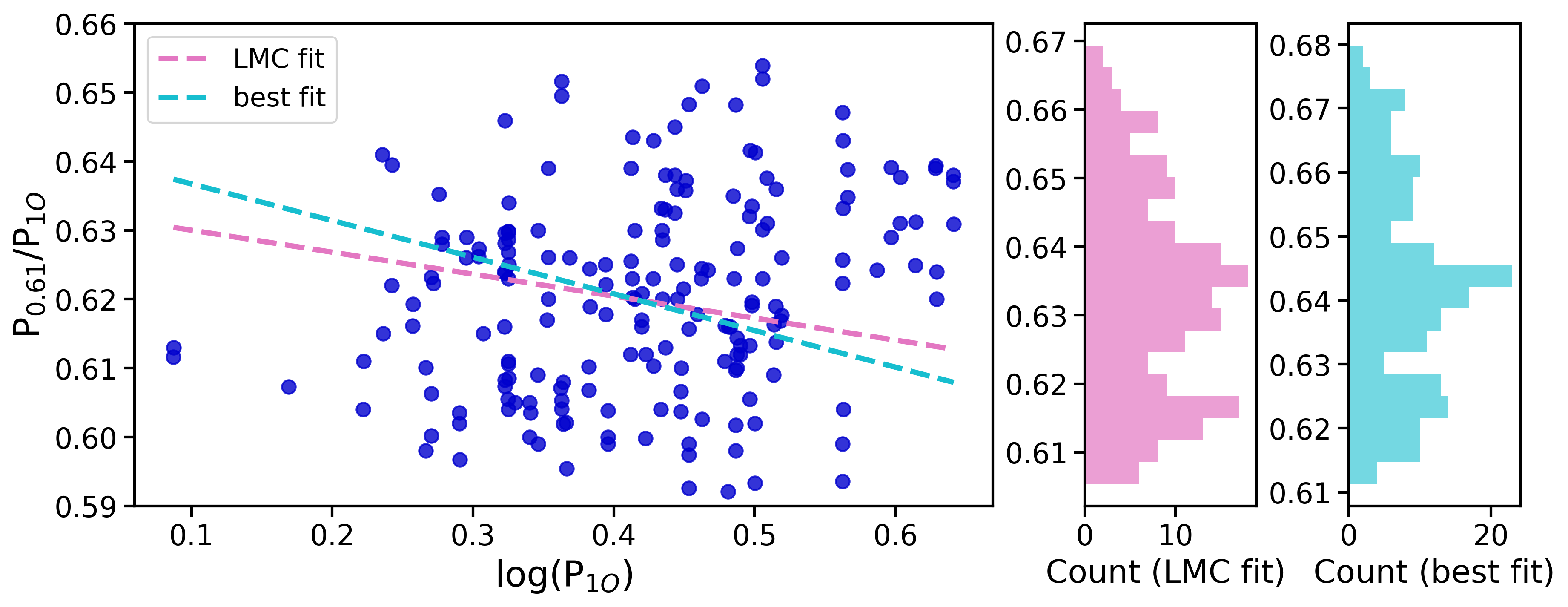}
\caption{Distributions of the projected period ratios (middle and right panels), perpendicular to the reference lines calculated for the middle sequence for LMC (pink) and to the best fit to the dots (cyan).}
\label{fig:sequences}
\end{figure}

The 0.61 group is displayed on the Petersen diagram in Fig. \ref{fig:petersen}. These are plotted together with the previously identified Galactic 0.61 stars. The values corresponding to the same stars are connected, and those that have the $f_{0.61}$ in addition are marked with a circle. The figure does not clearly show the three sequences, which are well-defined in the SMC and less pronounced in the LMC (see Fig.~5 in \citealt{smolec2023}). However, \citet{Rathour2021} have already shown some indications of a sequenced structure in the Galactic 0.61 group in the bottom panels of Fig. 6 in their paper. To check whether the sequences exist in the sample, we calculated projected distributions and displayed them in Fig.~\ref{fig:sequences}. First we used the slope that was determined for the middle sequence of the LMC stars ($P_{0.61}$/$P_{\rm{1O}}$ = a$\,$log$P_{\rm{1O}}$+$b$, where $a=-0.0319$, and $b=0.6332$ from \citealt{smolec2023}). This is marked with pink. The histogram shows that there is an indication of middle and bottom sequences, but existence of the upper sequence is ambiguous. We also  fitted the slope to the data, by optimizing the distribution to show more distinct groups. That resulted a steeper slope ($a=-0.0532$, and $b=0.6420$) shown in cyan. While the boundaries are slightly better visible, the upper sequence remains poorly defined.

In twelve stars, we were able to detect multiple $f_{0.61}$ peaks (indicated in red in Fig.~\ref{fig:petersen}). 
Six of these stars also exhibit multiple $f_{\rm{sh}}$ peaks, which were detected in either the same or in different sectors. A further fifteen stars only show multiplets at $f_{\rm{sh}}$. The vast majority of close frequencies  satisfy the traditional Rayleigh criterion for frequency resolution ($\Delta f = 1/{\rm T}$, where T is the length of the data). However, we identified frequency pairs closer than the Rayleigh frequency in thirteen stars of the 0.61 group. In five cases, these frequencies are clearly resolved in other sectors, suggesting that we can go slightly beyond 1/T. \cite{kallinger} have demonstrated on simulated data that peaks as close as $\sim$0.5/T can still be reliably distinguished. The closest frequencies in our sample are separated by 0.67/T (TIC 4712990), satisfying this criterion. 

\begin{figure}
\centering
\includegraphics[width=0.45\textwidth]{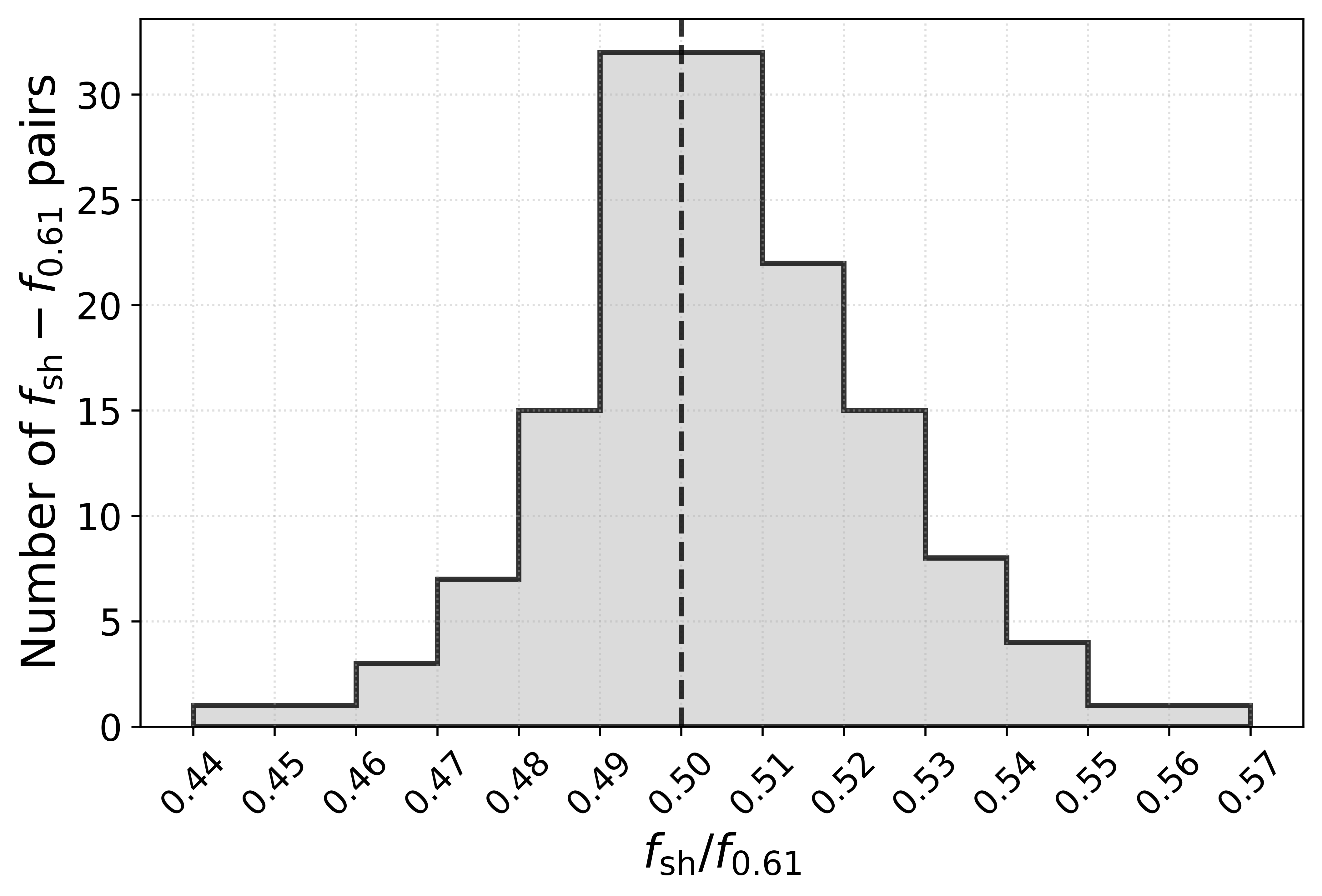}
\caption{Distribution of the $f_{\mathrm{sh}}/f_{\mathrm{0.61} } $ frequency ratios. Slight overabundance is visible on the right side.}
\label{fig:hist}
\end{figure}

The multiple $f_{0.61}$ peaks in Fig. \ref{fig:petersen} appear to belong to different sequences for some stars, while for others these peaks are too close to be identified as separate sequences. Additionally, the correspondence between the $f_{\rm{sh}}$ and the $f_{0.61}$ peaks remains ambiguous. We made an attempt to pair them as closely as possible to the 0.5 ratio. A more in-depth examination of the subharmonics reveals that the ratio can vary significantly, ranging from 0.44 to 0.56. As Fig.~\ref{fig:hist} illustrates, the distribution of these ratios is characterized by an observed asymmetry, which indicates that the  $f_{\rm{sh}}/f_{0.61}$>0.5 ratio is more prevalent (58 percent). Similar histograms were also produced for the Magellanic Clouds (see Fig.~7 in \citealt{smolec2023}), also finding the ratios above 0.5 to be slightly more frequent.

\begin{figure}
\centering
\includegraphics[width=0.49\textwidth]{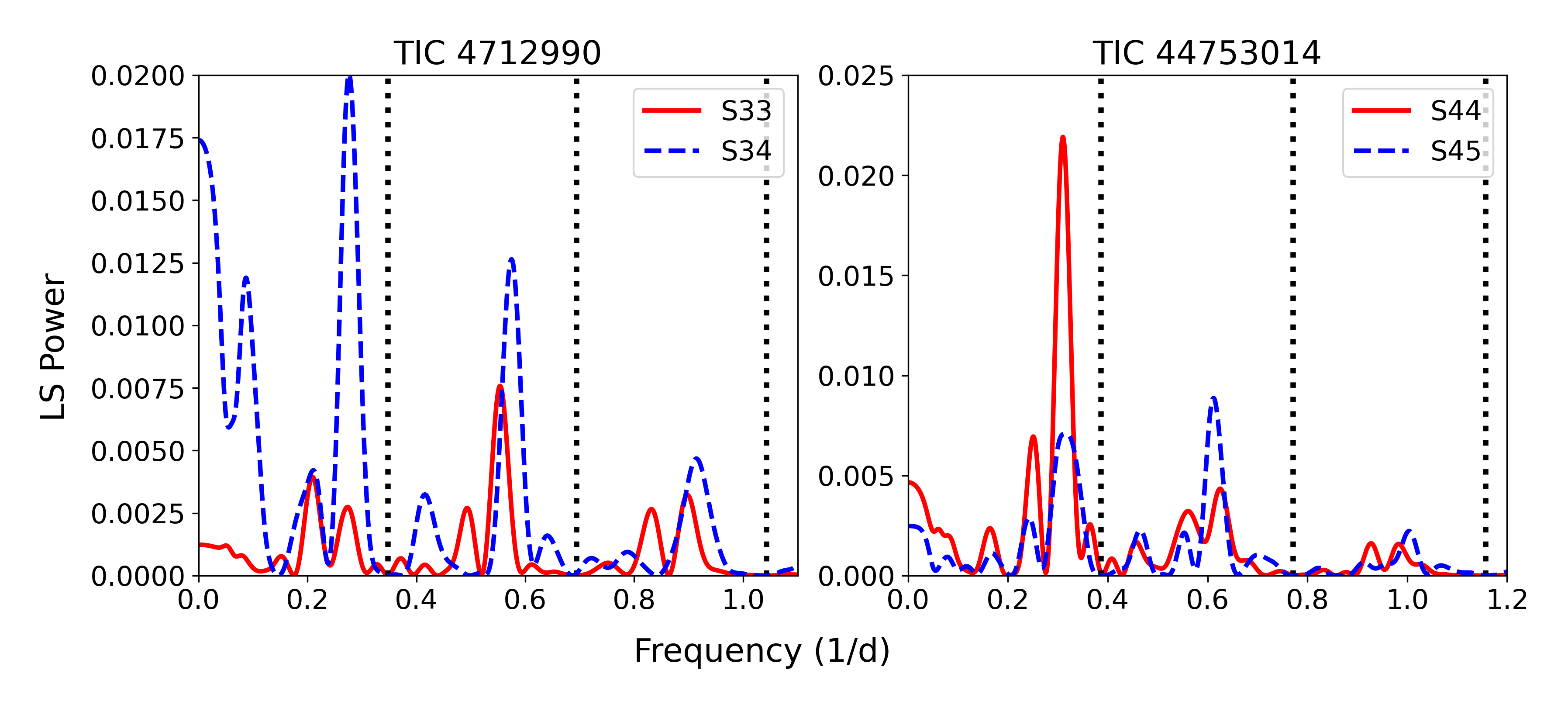}
\caption{Residual spectra for TIC 4712990 and TIC 44753014 after prewhitening with the main pulsational frequency and its harmonics ($f_{1\mathrm{O}}$, 2$f_{1\mathrm{O}}$, 3$f_{1\mathrm{O}}$ marked with dotted black lines) computed from data of consecutive sectors (red solid an blue dashed lines). The figures demonstrate amplitude variations of the $f_{\mathrm{sh}}$ and $f_{\mathrm{0.61}}$ detected in a short timescale.}
\label{fig:shvar}
\end{figure}

We observed that both $f_{0.61}$ and $f_{\rm{sh}}$ demonstrate significant amplitude variations over time. Which of the two frequencies is the dominant in the residual spectra often varies. This variation is even present between successive sectors, as demonstrated in Fig.~\ref{fig:shvar}. It is not clear whether this is caused by a real amplitude change or by the temporal appearance of nearby frequencies that are unresolved. The frequency values also often change between the different sectors, similar to what we saw at $f_{0.68}$. 
For 21 stars, we found that $f_{0.61}$ disappears in one or even several sectors, and only $f_{\rm{sh}}$ is detectable. We also found multiple subharmonics in some of the stars, some of which had no matching signal in the 0.6–0.65 period ratio (see Section \ref{sec:other}). We note here that the considerable number of stars with non-coherent $f_{0.61}$ and $f_{\rm{sh}}$ frequencies in the OGLE measurements reported by \citet{smolec2023} suggests that temporal variations among these signals are prevalent. Temporal variability of the 0.61 signal was also observed in RR Lyrae stars \citep{szabo-2014, Moskalik-2015,benko-2023, k2_rrc}.


\begin{figure*}
\centering
\includegraphics[width=0.95\textwidth]{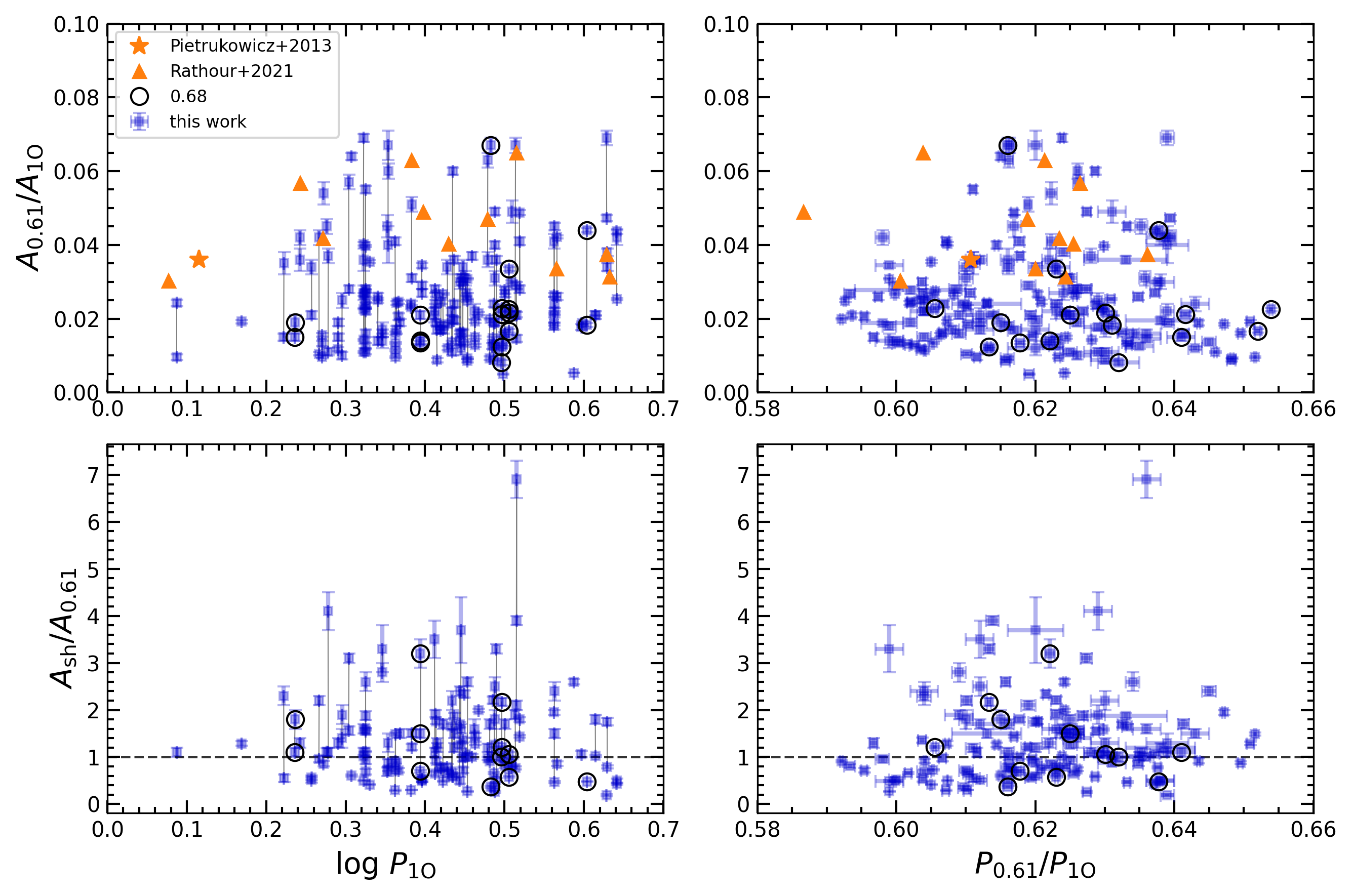}
\caption{Amplitude ratios of the 0.61 peaks relative to the first overtone mode, and the subharmonic peak. Left panels show the values as a function of main pulsation period, while right panels uses $P_{\mathrm{0.61} } / P_{1\mathrm{O}}$ period ratio. Symbols are the same as in Fig.~\ref{fig:petersen}. }
\label{fig:petersen4x}
\end{figure*}


We examined the amplitude ratios of the $f_{0.61}$ and $f_{\rm{sh}}$ peaks in more detail. In the upper panels of Fig.~\ref{fig:petersen4x}, the amplitude ratios of the 0.61 frequency with respect to the overtone mode are displayed. The values range from 0.005 to 0.07, with higher occurrences towards lower $A_{0.61}$/$A_{1\rm{O}}$ ratios. However, in the upper right panel, it is apparent that the higher amplitude ratio towards the central frequency ratios (i.e., the potential mid-sequence) is more frequent. 
If this is a real effect, then the limit of detectability is more likely to be reached in the middle sequence, which may result in a more populated appearance. The lower panels show the amplitude ratio of the subharmonics to the 0.61 peaks. Here, we find that the $A_{\rm{sh}}$ is more often greater than $A_{0.61}$ (in 58 percent of cases). With this feature, they resemble the LMC Cepheids, while in the SMC we observe the opposite. The presence of $f_{0.68}$ does not appear to have any effect on the amplitude ratios.

The 1.5$f_{0.61}$ is seen commonly in RR Lyrae stars, a frequency found in only nine of our stars. These are labelled 'hif' (half-integer frequencies) in the Table \ref{tab:061}. Interestingly, two of these stars (TIC 123734693 and TIC 234943347) also have the combination $f_{\rm{1O}}$+$1.5f_{0.61}$. In the former star, $1.5f_{0.61}$ is much stronger than $f_{0.61}$. We also searched for the combination of the subharmonic and the first overtone. In two stars, we found a signal that could match the combination of $f_{\rm{1O}}$+$f_{\rm{sh}}$ (TIC 126545806
and TIC 272999674). This combination is also observed in stars that are considered to be members of the 0.61 group, where only the subharmonic is detected (see section \ref{sec:other}). 

We identified the presence of quasi-symmetrically located peaks on either side of the overtone frequency in three stars. These could be signs of weak amplitude modulation, therefore we marked these stars as 'mod?' in the Table \ref{tab:061}. Further observations of these candidate modulated stars are necessary to confirm this suspicion.

In the 0.61 group we found four stars to be common with the findings of \citet{Rathour2021}: 
TIC 113843324, TIC 126545806, TIC 147184001 and TIC 317210786. 
A direct comparison of the $P_{0.61}/P_{\rm{1O}}$ ratios derived from TESS and OGLE data reveals discrepancies. However, this is to be expected from non-coherent signals investigated on very different timescales. Nevertheless, the fact that the 0.61 signal of these four stars has been detected by multiple surveys provides robust confirmation of their membership of the group. Moreover, the present analysis confirms the spectroscopic detection of the 0.61 signal in V0391 Nor (TIC 283374564) reported by \citet{veloce1}. Unfortunately, for the other stars in that study, BG Cru and QZ Nor, the 0.61 signal could not be detected unambiguously. 

For a significant proportion of the stars examined, we found that the signal was likely to be present, albeit not reaching a level of statistical significance or the residual spectrum was too blurred due to instrumental problems. These stars with uncertain detections are not listed here, but will be the subject of further future investigations, with the aim of solving the detrending problems mentioned in Section \ref{tessdata}. 

\subsubsection{Metallicity dependence of the 0.61 signal}

\begin{figure}
\centering
\includegraphics[width=0.49\textwidth]{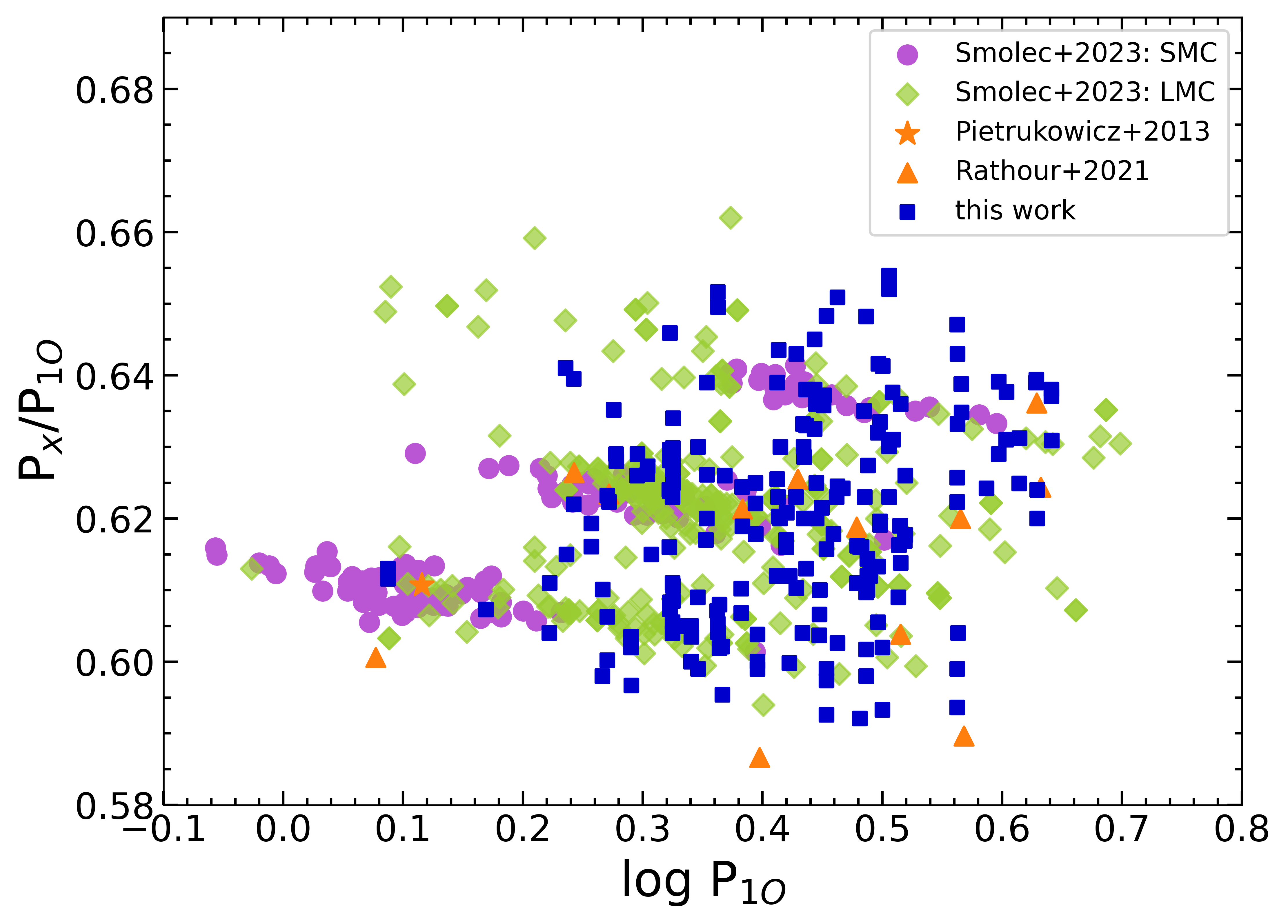}
\caption{Petersen diagram of 0.61 groups in the SMC (purple dots), LMC (green diamonds) and the Milky Way (orange and blue symbols) revealing a noticeable shift towards longer periods in the more metal-rich environments.}
\label{fig:shift}
\end{figure}

Metallicity may play an important role in the appearance of non-radial modes. \citet{Rathour2021} identified a systematic  shift towards longer periods with increasing metallicity when comparing the 0.61 sequences in the SMC, LMC, and the Milky Way. These sequences are illustrated in Figure \ref{fig:shift}, where we also included our results. The position of the Galactic sample in the Petersen diagram further supports the theory that the metallicity-related trend could be genuine. We calculated the mean periods to be 1.826, 2.375 and 2.685 days for the SMC, LMC and Galactic 0.61 stars, respectively. These values also indicate the trend, even though the longest-period stars could not be investigated with TESS.

\begin{figure}
\centering
\includegraphics[width=0.49\textwidth]{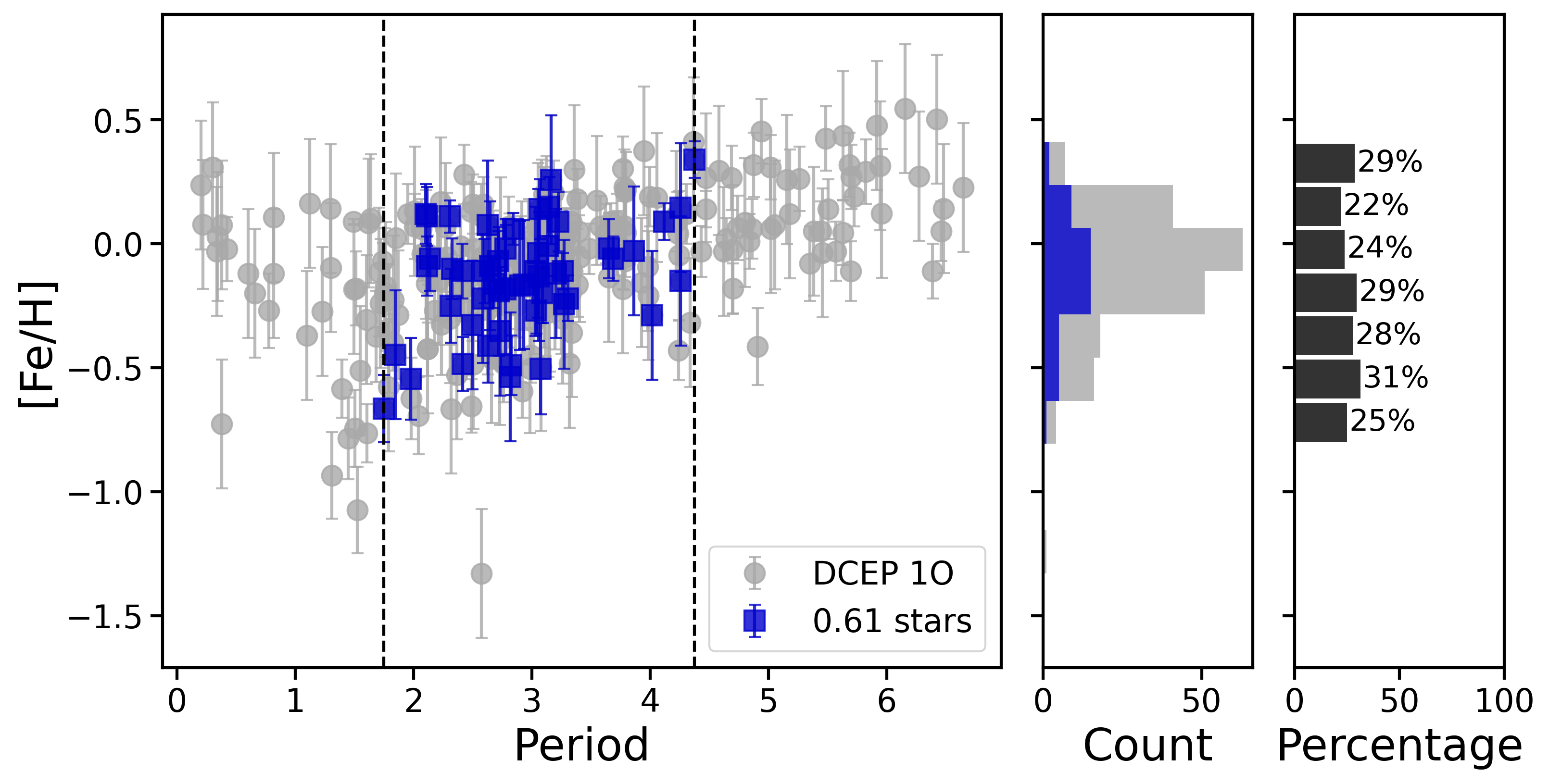}
\caption{Metallicity-period diagram of the 0.61 stars, based on data collected by \citet{trentin}. The metallicity distribution
is calculated for the period range within the dashed lines. No clear connection is visible between [Fe/H] and occurrence rates. 
}

\label{fig:metal}
\end{figure}

Theoretical calculations suggest that the incidence rate of non-radial modes decreases with decreasing metallicity for RR Lyrae models \citep{netzel-2024}. The distribution of photometric metallicity for TESS RRc stars with 0.61 signals, analyzed by \citet{benko-2023}, also support this prediction. With our newly discovered sample of 0.61 stars we can inspect whether similar connection exist in Cepheids. In this analysis, we used the recent collection of literature spectroscopic metallicities of Galactic Cepheids by \cite{trentin}. [Fe/H] indices are available for 282 first overtone stars in that catalog, of which 51 stars match with our stars in the 0.61 group. We could include one additional star from the 'Group1' by \citet{Rathour2021}, OGLE-GD-CEP-0041. The  values are displayed as a function of the pulsation period in Fig.~\ref{fig:metal}. The distribution was calculated for the period range in which the 0.61 stars were identified (indicated by the vertical dashed line). The histograms on the right side of the figure present the counts and percentages of the 0.61 stars in each bin. This latter indicates that the distribution is relatively uniform,  with no large variation observed against metallicity.
Nevertheless, the result must be treated with caution for two reasons. Firstly, the sample is small and it is possible that new discoveries will significantly change the picture. Secondly, the uncertainties of [Fe/H] indices are large, spanning up to a significant fraction of the parameter range.

\subsection{Other periodicities}
\label{sec:other}

\begin{figure}
\centering
\includegraphics[width=0.45\textwidth]{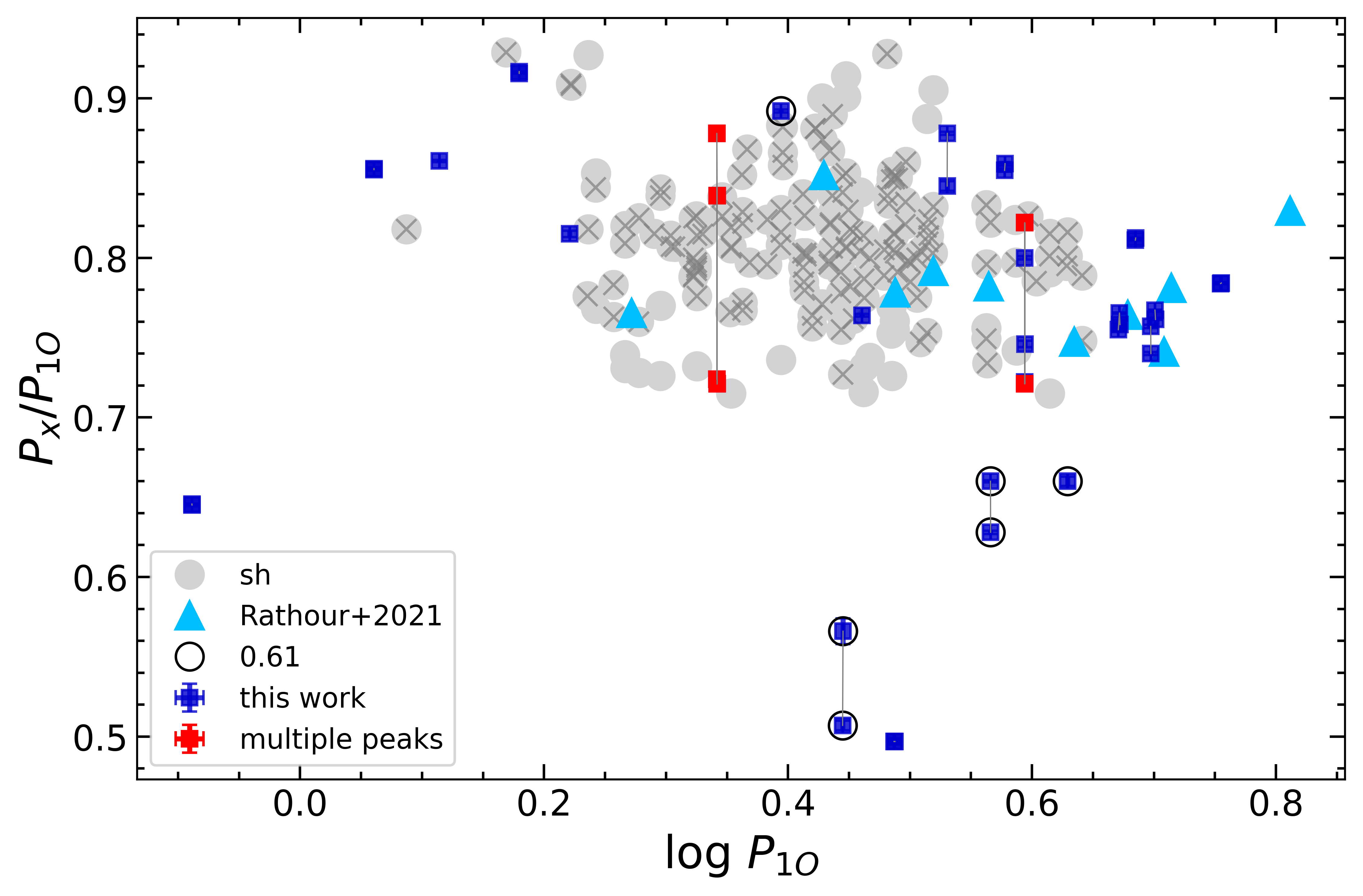}
\caption{Petersen diagram for the other additional modes. Stars labeled as 'Group 2' in the study of \citet{Rathour2021} are marked with light blue triangles. Grey dots indicate the positions of the subharmonics of the 0.61 group presented in Section \ref{sec:061}. The crosses denote those subharmonics for which a matching $f_{0.61}$ peak was identified.}

\label{fig:otheradd}
\end{figure}

We found several stars with low-amplitude periodicities that fall outside of the 0.68 and 0.61 periodicity ranges. These are listed in Table \ref{tab:add}. The vast majority of these periodicities have periods longer than the overtone mode. We found signals with shorter periodicities only in three stars. Of these, we observe the combination peak with the overtone in one case (TIC~281312636). Signals that have a period longer than the overtone are shown in Fig.~\ref{fig:otheradd}.

Some of these additional signals fall within the range in which we would expect to find the period ratios of subharmonic frequencies to the first overtone mode. So, it can be hypothesized that these are members of the 0.61 group, wherein solely the subharmonic, i.e., the theoretical non-radial mode frequencies are observed. In Fig.~\ref{fig:otheradd}, the period ratios between the subharmonics and the first overtone mode from the 0.61 group are shown with grey symbols. The periodicities identified in this range by \citet{Rathour2021} and designated as 'Group 2' are indicated by light blue triangles. They also speculated that these could be direct detections of non-radial modes. The same hypothesis was also mentioned in the study of the Magellanic Clouds for signals in the same range by \citet{smolec2023}. In a subset of these cases, the $f_{\rm{1O}}+f_{\rm{x}}$ combination frequency also appears. The combination frequencies of the subharmonic with the first overtone mode have previously been simulated for RRc stars \citep{netzel-kolenberg} and have been found in observations \citep{benko-2023}. 

In Fig.~\ref{fig:otheradd}, the presence  of a matching $f_{0.61}$ peak for the subharmonics is indicated by crosses.
There are many subharmonics that have no matching $f_{0.61}$ peak, as shown by the grey dots without crosses found at both the lower and higher edges of the subharmonic range.

Signals also appear far from the subharmonic cluster, even at a period ratio of $\sim$0.5. For example in TIC~39141607, of which this strange periodicity can also be seen in the photometric data collected by the All Sky Automated Survey for SuperNovae\footnote{\url{https://asas-sn.osu.edu/variables/}} (ASAS-SN, \citealt{asas-sn}). Additional frequency peaks found at a period ratio higher than 0.95 with the main pulsation mode were not included here, as we cannot distinguish them from instrumental signals (e.g., peaks due to amplitude differences between orbits and sectors). For the same reasons we ignored the very low frequency signals as well. Moreover, TESS measurements have a relatively poor spatial resolution, and there is a high chance of contamination. So, we are unsure of the origin of additional low-amplitude signals outside the known groups, especially if they have no combination frequency with the overtone mode.

\begin{figure}
\centering
\includegraphics[width=0.49\textwidth]{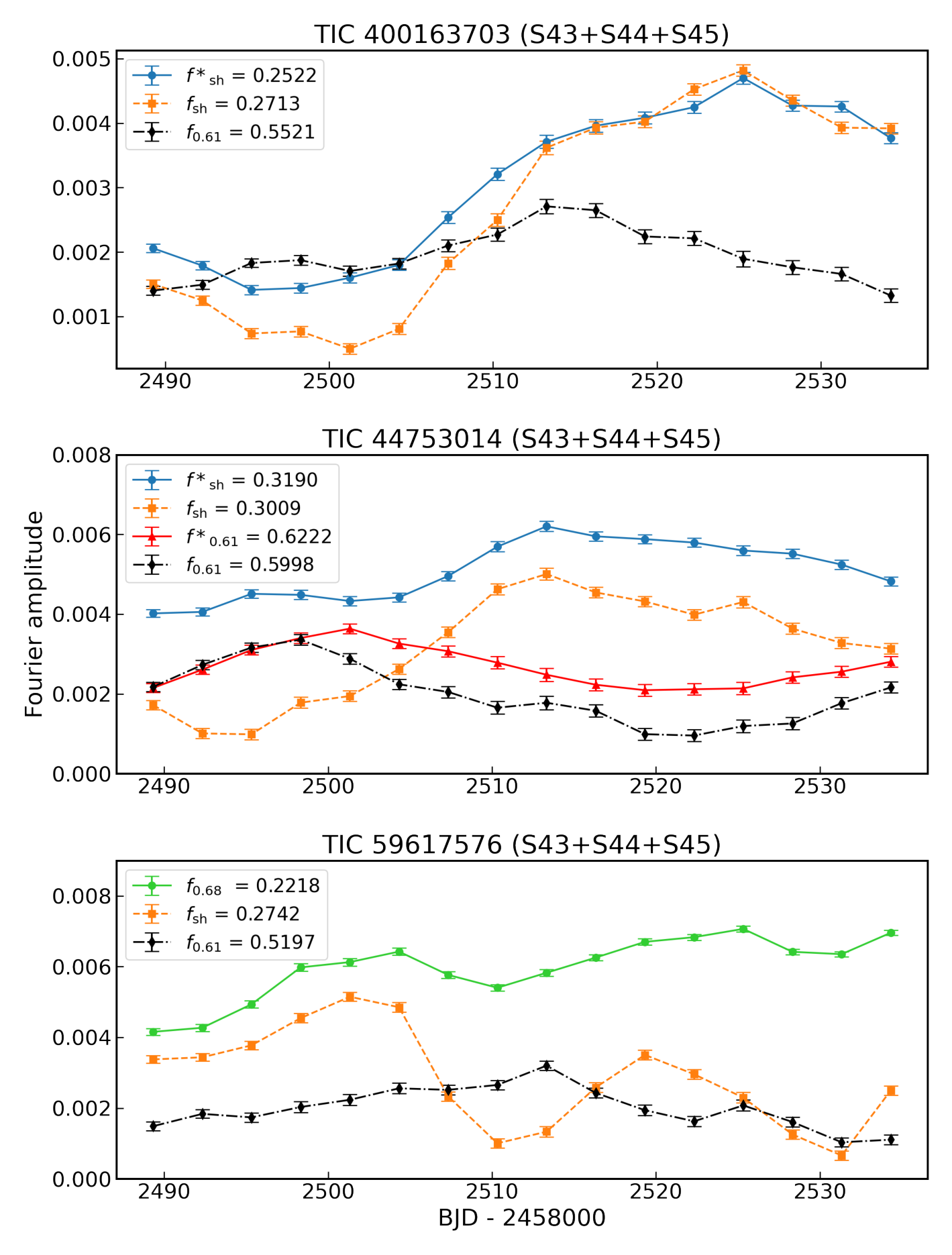}
\caption{The amplitude evolution of the low amplitude signals of TIC~400163703, TIC~44753014 and TIC~59617576 observed in three consecutive sectors. The asterisk indicates the higher amplitude peak of the subharmonics and the associated 0.61 frequency where it is present.
}

\label{fig:ampevol}
\end{figure}

\subsection{Comments on temporal variation and stars with no additional frequencies}

As demonstrated in Fig.~\ref{fig:shvar} temporal variation of the additional signals may occur on a timescale as short as one month. 
To conduct a more detailed investigation, we used time-dependent analysis with a 30-day sliding window with 3-day step size for stars observed over three consecutive sectors. The amplitude variations of additional frequencies in three star are presented in Fig.~\ref{fig:ampevol}. Despite the limited timespan, which prevents observation of cyclic variations, it is possible to see that the change in the amplitude of the 0.61 signal and its subharmonic is different, and sometimes they appear to be anti-correlated. The amplitude variation of $f_{0.68}$ in the third star also does not follow any of the other two signals. In the case of other stars within the 0.61 group, it is seen that an increase in the amplitude of  $f_{\rm{sh}}$ is accompanied by the vanishing of $f_{0.61}$, and the combination of $f_{\rm{1O}}$+$f_{\rm{sh}}$ becomes visible. We found no examples in which the 0.61 signal and subharmonic both completely disappear at the same time.

The underlying physics of the temporal variation of these signals is an open question and would require detailed modelling to understand. In Dziembowski's model of a slowly rotating and oscillating star, geometric cancellation plays a crucial role in the appearance of $f_{0.61}$ and $f_{\rm{sh}}$. It has been shown that a much simpler frequency spectrum and less variability are expected near $f_{0.61}$ than at $f_{\rm{sh}}$ (see Section 3 in \citealt{dziem}). This is consistent with what we observe in Fig.~\ref{fig:ampevol}. We may speculate that an oblique pulsation axis could also be behind this variability, in which case a cyclic variation with the rotation period would be expected. The estimated rotation periods for Classical Cepheids are consistent with this hypothesis (see \citealt{Anderson-2016}). On the other hand, if the visibility of the pulsation geometry is unchanged, we could observe the beating of a rotational multiplet (e.g., $m=-1,0,+1$), which might appear as a complex, quasi-periodic change. The 0.61 signal and subharmonics have also been found to be highly variable in RRc stars in the data of various space observations
\citep{szabo-2014, Moskalik-2015, k2_rrc, benko-2023}. However, the variability occurs on a time scale ranging from 10 to 200 days and is characterized by its irregular nature. It is therefore evident that rotation cannot play a role in this variability, since RR~Lyrae stars are much slower rotators \citep{peterson-1996}.

\begin{figure}
\centering
\includegraphics[width=0.49\textwidth]{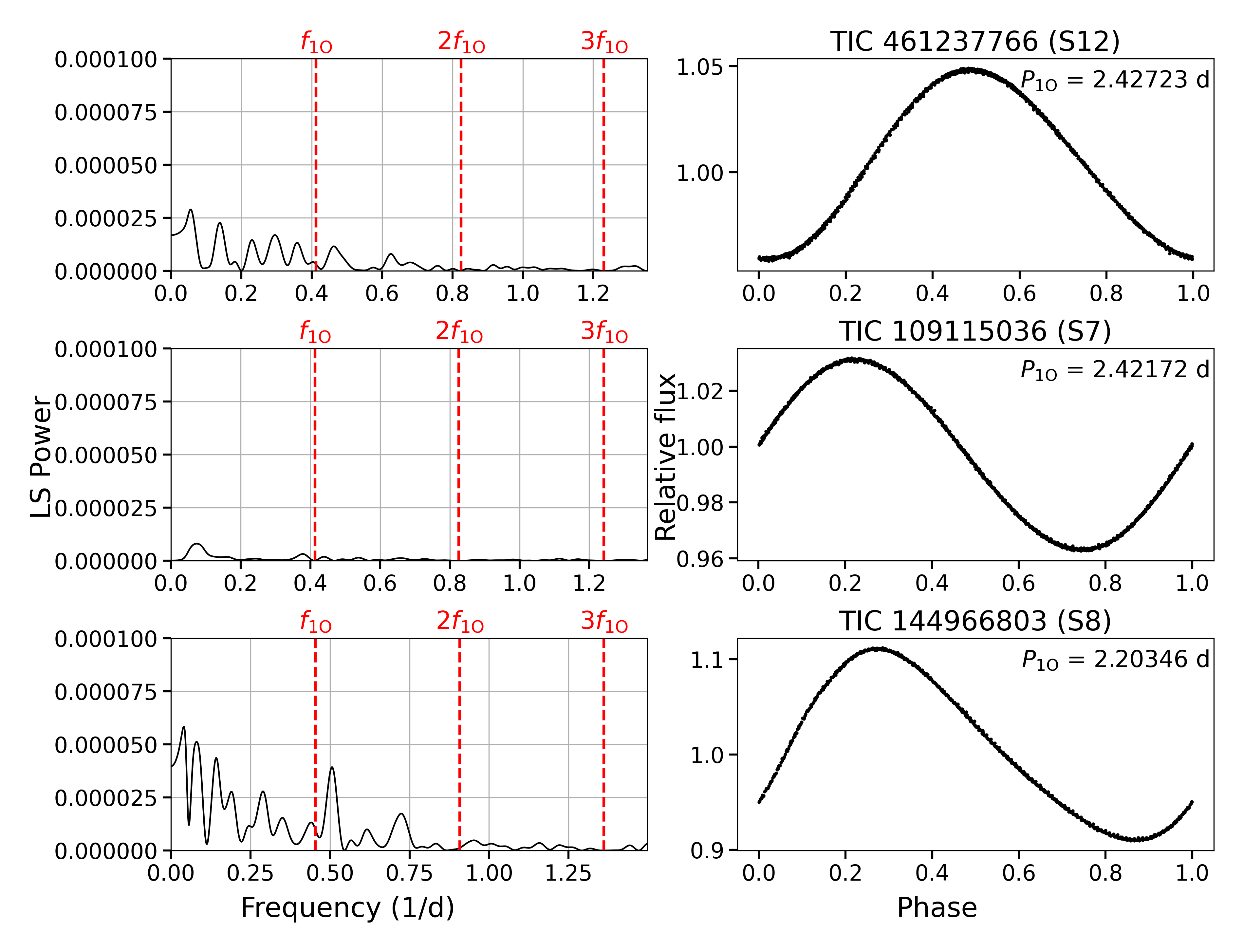}
\caption{Residual spectra and phase folded light curves of example first overtone Cepheids with no additional signals.}

\label{fig:nofx}
\end{figure}

It is important to note that the analysis presented is very sensitive to the quality of the data. We tried to use as many TESS sectors as possible for each star, but not all of them were suitable for unambiguous analysis. These sectors are not included in the tables of the Appendix. On the other hand, we have found several stars with high quality data in which no additional signal can be found. We illustrate this in Fig.~\ref{fig:nofx} where we plot the residual spectra and the phase folded light curves in three stars that clearly show the absence of additional signals.

\section{Summary}

The majority of Galactic classical Cepheids have been observed by TESS. However, the mission's design and the quality issues allow us to conduct a search for low-amplitude periodicities only on a limited sample size. A brief summary of our results is presented below.

 \begin{itemize}
     
    \item We identified a total of 17 radial double mode (1O2O) stars, including confirmation of two that had previously been suspected of belonging to the class. Furthermore, we observed large changes in mode amplitudes between the sectors from different years, which may be an indication of a periodic amplitude modulation phenomenon.
    
    \item We discovered 79 new members of the 0.61 group and 14 new members of the 0.68 group of overtone Cepheid stars. Six of these show both signals. Additionally, we confirmed four of the former group and one of the latter group that had previously been discovered. The sequences in the Petersen diagram of the 0.61 are significantly less clear than those observed for SMC or LMC. Of the three sequences, only two are discernible.
       
    \item We observed temporal variations in frequencies and amplitudes in both the 0.61 and 0.68 signals. The timescale appears to be on the order of months for some stars. The origin of the variation remains unclear. In several cases, amplitude modulation of the radial mode is also suspected. Longer, uninterrupted high-cadence space observation may require a more detailed study of temporal changes.

    \item We found additional low-amplitude frequencies outside the period ranges of the second overtone mode and the 0.68 and 0.61 signals. In 11 stars, these frequencies are close to the range where subharmonic frequencies, i.e., the signal of the non-radial mode, are expected. We suggest that these stars are candidate 0.61 stars.

    
\end{itemize}

These newly discovered 0.61 stars, added to the previous discoveries of \citet{Rathour2021} and \citet{veloce1}, give us an incidence rate lower limit of $\sim$8.7 percent of all overtone Cepheids in the Milky Way. Taking into account the 13.5 $T_{\rm{mag}}$ brightness limit we applied, this rises to around 18 percent. For comparison it was found to be 12.6 and 16.5 percent in the SMC and LMC, respectively (see \citealt{smolec2023}). We suspect that there are many more stars with low-amplitude periodicities that we were unable to reveal due to instrumental trends. It is challenging to distinguish between intrinsic and instrumental additional variability in data with a low number of cycles. Developing a detrending algorithm to resolve this problem could significantly improve detection. It is also expected that future TESS observations of regions in the Galactic Plane will yield further insights into the pulsation nature of Classical Cepheids.

\begin{acknowledgements}

This paper includes data collected by the and \textit{TESS} mission. Funding for the missions are provided by the NASA Science Mission Directorate. The research was partially supported by the ‘SeismoLab’ KKP-137523 \'Elvonal grant of the Hungarian Research, Development and Innovation Office (NKFIH). Some \texttt{python} codes were developed with the help of ChatGPT 4.0. The authors would also
like to gratefully thank J\'ozsef Benk{\H{o}} for the fruitful discussions on this topic. HN acknowledges support from the European Research Council (ERC) under the European Union’s Horizon 2020 research and innovation program (grant agreement No. 951549 - UniverScale). We thank the referee for their comments that helped to improve the paper.

\end{acknowledgements}

%
%

\bibliographystyle{aa} 
\bibliography{TESSO1_clean} 

\onecolumn

\appendix
\section{List of stars with additional periodicities}

\begin{table*}[ht!]

\caption{New 1O2O stars found in the analyzed sample. Fourier frequencies, amplitudes and ratios of the stars with additional second overtone radial mode derived from data of multiple sectors.}

    \centering
    \begin{tabular}{llllllll}
    
TIC &  Sectors & $f_{\rm{1O}}$ & $A_{\rm{1O}}$ & $f_{\rm{2O}}$  &  $A_{\rm{2O}}$ &  $f_{\rm{1O}}$/ f$_{\rm{2O}}$&  $A_{\rm{2O}}$/$A_{\rm{1O}}$\\

\noalign{\vskip 2pt} 
\hline
\hline
\noalign{\vskip 2pt} 

64287862	 & 	S15+S16	 & 	2.2947(2)	 & 	0.0597(8)	 & 	2.8675(2)	 & 	0.0239(3)	 & 	0.80026(7)	 & 	0.401(8)\\
	 & 	S56+S57	 & 	2.29508(3)	 & 	0.0640(2)	 & 	2.86753(6)	 & 	0.01455(9)	 & 	0.80037(2)	 & 	0.228(2)\\
	 & 	S76+S77	 & 	2.29486(4)	 & 	0.0662(3)	 & 	2.86746(5)	 & 	0.0248(1)	 & 	0.80031(2)	 & 	0.375(2)\\
75696685	 & 	S43+S44+S45	 & 	2.65595(3)	 & 	0.1219(5)	 & 	3.32335(7)	 & 	0.0232(2)	 & 	0.79918(2)	 & 	0.191(2)\\
	 & 	S71+S72	 & 	2.65603(3)	 & 	0.1186(3)	 & 	3.32384(9)	 & 	0.0130(1)	 & 	0.79909(2)	 & 	0.1098(9)\\
78429604	 & 	S43+S44+S45	 & 	2.68804(3)	 & 	0.0988(4)	 & 	3.3533(1)	 & 	0.0102(2)	 & 	0.80162(3)	 & 	0.104(2)\\
	 & 	S72	 & 	2.6879(1)	 & 	0.0989(4)	 & 	3.3537(3)	 & 	0.0188(2)	 & 	0.80148(7)	 & 	0.190(2)\\
91460548	 & 	S14	 & 	2.6562(1)	 & 	0.0835(6)	 & 	3.3199(6)	 & 	0.0061(2)	 & 	0.8001(1)	 & 	0.073(2)\\
	 & 	S41	 & 	2.65629(8)	 & 	0.0822(3)	 & 	3.3199(3)	 & 	0.0094(1)	 & 	0.80010(6)	 & 	0.114(1)\\
	 & 	S54+S55	 & 	2.65631(3)	 & 	0.0745(2)	 & 	3.32019(8)	 & 	0.00974(8)	 & 	0.80005(2)	 & 	0.131(1)\\
	 & 	S75	 & 	2.65635(4)	 & 	0.0707(1)	 & 	3.3202(1)	 & 	0.00915(6)	 & 	0.80006(3)	 & 	0.1295(9)\\
	 & 	S81	 & 	2.65640(5)	 & 	0.0679(2)	 & 	3.3204(1)	 & 	0.00873(6)	 & 	0.80002(4)	 & 	0.1285(9)\\
184302985	 & 	S14	 & 	2.9785(3)	 & 	0.088(1)	 & 	3.7277(5)	 & 	0.0171(4)	 & 	0.7990(1)	 & 	0.194(5)\\
 & 	S41	 & 	2.9787(1)	 & 	0.0874(5)	 & 	3.7281(2)	 & 	0.0138(2)	 & 	0.79898(6)	 & 	0.158(2)\\
	 & 	S54+S55	 & 	2.97856(5)	 & 	0.0892(4)	 & 	3.72788(8)	 & 	0.0222(2)	 & 	0.79900(2)	 & 	0.249(2)\\
	 & 	S74+S75	 & 	2.97849(2)	 & 	0.0867(2)	 & 	3.72800(6)	 & 	0.01179(7)	 & 	0.79895(1)	 & 	0.1361(8)\\
	 & 	S81	 & 	2.97852(6)	 & 	0.0890(3)	 & 	3.7280(2)	 & 	0.0144(1)	 & 	0.79895(4)	 & 	0.162(1)\\
192964026	 & 	S39	 & 	3.3788(5)	 & 	0.048(1)	 & 	4.234(1)	 & 	0.017(1)	 & 	0.7980(3)	 & 	0.34(2)\\
195587833	 & 	S15	 & 	1.9738(3)	 & 	0.098(1)	 & 	2.4627(8)	 & 	0.0147(4)	 & 	0.8015(3)	 & 	0.150(5)\\
	 & 	S55+S56	 & 	1.97387(3)	 & 	0.0934(3)	 & 	2.46257(8)	 & 	0.01100(9)	 & 	0.80155(3)	 & 	0.118(1)\\
	 & 	S75+S76	 & 	1.97381(3)	 & 	0.0901(2)	 & 	2.46225(5)	 & 	0.0209(1)	 & 	0.80163(2)	 & 	0.232(1)\\
263385369	 & 	S12	 & 	1.6712(4)	 & 	0.116(3)	 & 	2.085(3)	 & 	0.015(2)	 & 	0.802(1)	 & 	0.13(2)\\
	 & 	S39	 & 	1.6714(2)	 & 	0.114(1)	 & 	2.088(1)	 & 	0.025(1)	 & 	0.8006(4)	 & 	0.22(1)\\
	 & 	S65	 & 	1.6714(1)	 & 	0.1311(8)	 & 	2.0857(5)	 & 	0.0299(6)	 & 	0.8014(2)	 & 	0.228(5)\\
290277380	 & 	S15+S16	 & 	3.74071(8)	 & 	0.0925(7)	 & 	4.6773(2)	 & 	0.0101(2)	 & 	0.79975(4)	 & 	0.109(2)\\
	 & 	S55+S56	 & 	3.74113(2)	 & 	0.0965(2)	 & 	4.67789(5)	 & 	0.01267(6)	 & 	0.79975(1)	 & 	0.1313(7)\\
	 & 	S75+S76	 & 	3.74099(2)	 & 	0.0944(2)	 & 	4.67772(5)	 & 	0.01056(5)	 & 	0.799746(9)	 & 	0.1118(6)\\
316952151	 & 	S18+S19	 & 	1.8822(1)	 & 	0.0692(7)	 & 	2.3466(2)	 & 	0.0183(3)	 & 	0.80208(8)	 & 	0.265(5)\\
	 & 	S59	 & 	1.8825(1)	 & 	0.0640(3)	 & 	2.3466(2)	 & 	0.0156(2)	 & 	0.80222(9)	 & 	0.243(3)\\
324664178	 & 	S10+S11	 & 	1.91699(9)	 & 	0.0691(6)	 & 	2.3915(3)	 & 	0.0108(3)	 & 	0.80157(9)	 & 	0.156(4)\\
	 & 	S37+S38	 & 	1.91707(4)	 & 	0.0865(4)	 & 	2.3922(2)	 & 	0.0098(2)	 & 	0.80138(6)	 & 	0.113(2)\\
	 & 	S64	 & 	1.91706(7)	 & 	0.0948(3)	 & 	2.3923(4)	 & 	0.0076(2)	 & 	0.8014(1)	 & 	0.080(2)\\
363328038	& S9	& 4.0324(2)	& 0.0494(4) &	5.052(1)	&0.0046(3)&	0.7982(2)&	0.094(6) \\
364898211	 & 	S10+S11	 & 	0.71873(5)	 & 	0.0929(4)	 & 	0.8935(1)	 & 	0.0088(1)	 & 	0.8044(1)	 & 	0.094(1)\\
	 & 	S37+S38	 & 	0.71865(7)	 & 	0.1093(7)	 & 	0.89365(5)	 & 	0.0507(2)	 & 	0.80417(9)	 & 	0.464(4)\\
	 & 	S64+S65	 & 	0.71860(2)	 & 	0.1476(3)	 & 	0.89319(8)	 & 	0.00658(5)	 & 	0.80453(8)	 & 	0.0446(4)\\
430122360	 & 	S16+S17	 & 	1.3194(1)	 & 	0.0904(9)	 & 	1.6381(2)	 & 	0.0221(4)	 & 	0.8055(1)	 & 	0.244(5)\\
	 & 	S56+S57	 & 	1.31921(3)	 & 	0.0996(3)	 & 	1.63820(5)	 & 	0.0267(1)	 & 	0.80528(3)	 & 	0.268(2)\\
	 & 	S76+S77	 & 	1.31916(3)	 & 	0.0949(3)	 & 	1.63794(5)	 & 	0.0250(1)	 & 	0.80537(3)	 & 	0.263(2)\\
438112973	 & 	S6	 & 	2.5606(4)	 & 	0.101(2)	 & 	3.1988(6)	 & 	0.0322(8)	 & 	0.8005(2)	 & 	0.319(9)\\
	 & 	S33	 & 	2.5606(1)	 & 	0.1229(7)	 & 	3.196(2)	 & 	0.0034(3)	 & 	0.8012(4)	 & 	0.028(2)\\
	 & 	S43+S44+S45	 & 	2.56057(2)	 & 	0.1178(4)	 & 	3.1982(1)	 & 	0.0062(1)	 & 	0.80062(3)	 & 	0.0529(9)\\
	 & 	S71	 & 	2.5606(1)	 & 	0.1133(5)	 & 	3.1983(2)	 & 	0.0288(2)	 & 	0.80062(6)	 & 	0.255(2)\\
442907962	 & 	S9	 & 	0.89059(8)	 & 	0.0605(4)	 & 	1.1055(6)	 & 	0.0057(3)	 & 	0.8056(5)	 & 	0.094(6)\\
455190351	 & 	S12	 & 	2.4545(3)	 & 	0.0618(9)	 & 	3.0852(3)	 & 	0.0237(3)	 & 	0.7956(1)	 & 	0.383(8)\\
	 & 	S38+S39	 & 	2.45463(2)	 & 	0.0634(2)	 & 	3.086(1)	 & 	0.00073(7)	 & 	0.7953(3)	 & 	0.012(1)\\
	 & 	S65	 & 	2.45458(5)	 & 	0.0689(1)	 & 	3.0848(3)	 & 	0.00504(7)	 & 	0.79570(8)	 & 	0.073(1)\\

     \hline

    \end{tabular}
         
    \label{tab:dm}
   
\end{table*}

\begin{table*}
    \caption{Stars with additional frequencies near the $P_{\rm{1O}}$/$P_{\rm{0.68}}$$\sim$0.68 ratio.    (c -- combination frequency, af -- additional frequency, 0.61 -- additional signal in the $P_{\rm{0.61}}$/$P_{\rm{1O}}$: 0.60--0.65 range.) }
    \centering
    \begin{tabular}{llllllll}

TIC &  Sectors & $f_{\rm{1O}}$ & $A_{\rm{1O}}$ & $f_{\rm{0.68}}$  &  $f_{\rm{0.68}}$/ $f_{\rm{1O}}$&  $A_{\rm{0.68}}$/$A_{\rm{1O}}$& Remarks\\

\noalign{\vskip 2pt} 
\hline
\hline
\noalign{\vskip 2pt} 

28248697	&	S19	&	0.4390(1)	&	0.0567(2)	&	0.2991(8)	&	0.681(2)	&	0.044(2)	&	--	\\
	&	S59	&	0.43918(3)	&	0.05763(8)	&	0.2982(4)	&	0.6790(8)	&	0.0419(7)	&	--	\\
52356596	&	S18	&	0.6068(1)	&	0.0450(2)	&	0.419(1)	&	0.690(2)	&	0.039(2)	&	--	\\
	&	S24+S25	&	0.60693(3)	&	0.0459(2)	&	0.4153(4)	&	0.6842(7)	&	0.035(1)	&	--	\\
	&	S58	&	0.60716(4)	&	0.0548(1)	&	0.4240(6)	&	0.698(1)	&	0.038(1)	&	--	\\
59617576	&	S6	&	0.31904(9)	&	0.0983(3)	&	0.2151(5)	&	0.674(2)	&	0.0428(8)	&	f$_{0.61}$	\\
	&	S33	&	0.31864(4)	&	0.0979(2)	&	0.2207(3)	&	0.6925(9)	&	0.0503(7)	&	f$_{0.61}$	\\
	&	S43+S44+S45	&	0.318754(9)	&	0.1010(1)	&	0.22179(8)	&	0.6958(2)	&	0.0560(6)	&	f$_{0.61}$	\\
	&	S71+S72	&	0.31884(1)	&	0.10049(9)	&	0.22159(8)	&	0.6950(2)	&	0.0598(4)	&	f$_{0.61}$	\\
81352154	&	S7	&	0.2365(1)	&	0.1199(8)	&	0.1614(7)	&	0.683(3)	&	0.033(1)	&	--	\\
	&	S34+S35	&	0.23652(3)	&	0.1117(3)	&	0.1641(2)	&	0.6937(8)	&	0.0301(5)	&	--	\\
	&	S61+S62	&	0.23650(2)	&	0.1060(2)	&	0.16195(7)	&	0.6848(3)	&	0.0319(2)	&	--	\\
89286690	&	S6	&	0.5479(1)	&	0.0533(2)	&	0.371(2)	&	0.676(4)	&	0.022(2)	&	--	\\
	&	S33	&	0.54796(5)	&	0.0624(1)	&	0.374(1)	&	0.683(2)	&	0.025(1)	&	--	\\
	&	S43+S44+S45	&	0.54791(1)	&	0.06327(9)	&	0.3747(3)	&	0.6838(5)	&	0.0249(9)	&	--	\\
	&	S71+S72	&	0.54797(1)	&	0.06746(8)	&	0.3736(4)	&	0.6817(7)	&	0.0235(8)	&	--	\\
197776441	&	S16+S17	&	0.34396(3)	&	0.06620(9)	&	0.2389(4)	&	0.695(1)	&	0.0324(5)	&	--	\\
	&	S56+S57	&	0.343675(4)	&	0.06283(3)	&	0.23634(7)	&	0.6877(2)	&	0.0329(2)	&	--	\\
	&	S76+S77	&	0.343876(4)	&	0.06386(3)	&	0.23571(5)	&	0.6855(1)	&	0.0373(2)	&	--	\\
235352769	&	S6	&	1.2866(2)	&	0.0867(7)	&	0.907(5)	&	0.705(4)	&	0.009(2)	&	--	\\
	&	S33	&	1.28637(9)	&	0.0952(4)	&	0.909(2)	&	0.707(1)	&	0.0124(9)	&	--	\\
239378995	&	S55+S56	&	0.22368(2)	&	0.0593(1)	&	0.15576(9)	&	0.6963(4)	&	0.0735(7)	&	c, af	\\
	&	S75+S76	&	0.22366(1)	&	0.05964(9)	&	0.15546(8)	&	0.6951(4)	&	0.0796(7)	&	af	\\
251182052	&	S33	&	0.32906(7)	&	0.1305(4)	&	0.2234(5)	&	0.679(1)	&	0.066(2)	&	f$_{0.61}$	\\
255977934	&	S18	&	0.40364(8)	&	0.0862(3)	&	0.280(1)	&	0.692(3)	&	0.043(2)	&	f$_{0.61}$	\\
	&	S24+S25	&	0.40345(2)	&	0.0850(2)	&	0.2754(6)	&	0.683(2)	&	0.021(1)	&	f$_{0.61}$	\\
	&	S58	&	0.40343(1)	&	0.08538(6)	&	0.2779(6)	&	0.689(2)	&	0.0138(4)	&	f$_{0.61}$, af	\\
307298395	&	S7	&	1.5189(2)	&	0.0690(5)	&	1.021(2)	&	0.672(1)	&	0.019(1)	&	--	\\
	&	S33+S34	&	1.51889(3)	&	0.0786(2)	&	1.0211(5)	&	0.6723(3)	&	0.0175(8)	&	c	\\
347692492	&	S58	&	0.56341(2)	&	0.06732(8)	&	0.3832(3)	&	0.6801(5)	&	0.0186(3)	&	c	\\
365297180	&	S15+S16	&	0.31214(3)	&	0.0991(3)	&	0.2114(4)	&	0.677(1)	&	0.044(2)	&	--	\\
	&	S56+S57	&	0.312177(7)	&	0.09536(7)	&	0.2174(1)	&	0.6963(4)	&	0.0416(6)	&	f$_{0.61}$	\\
	&	S76+S77	&	0.312075(8)	&	0.09989(8)	&	0.2181(1)	&	0.6989(4)	&	0.0453(6)	&	f$_{0.61}$	\\
395471265	&	S54	&	0.24922(5)	&	0.0719(2)	&	0.1690(2)	&	0.678(1)	&	0.0781(9)	&	f$_{0.61}$	\\
	&	S80+S81	&	0.24905(2)	&	0.0730(1)	&	0.1680(2)	&	0.6748(8)	&	0.059(1)	&	f$_{0.61}$	\\
722289976	&	S44	&	0.58120(9)	&	0.1167(4)	&	0.395(1)	&	0.679(6)	&	0.030(2)	&	f$_{0.61}$	\\   
	&	S45	&	0.58035(8)	&	0.1211(4)	&	0.400(2)	&	0.688(4)	&	0.015(1)	&	f$_{0.61}$	\\

     \hline

    \end{tabular}
         
    \label{tab:068}
   
\end{table*}

\clearpage
\setlength{\tabcolsep}{4pt}

\begin{longtable}{llllllllll}


\caption{%
Stars with additional frequencies near the $P_{\rm{0.61}}$/$P_{\rm{1O}}$$\sim$0.60--0.65 ratio.    (c -- combination frequency, hif -- half-integer frequency at 1.5~f$_{0.61}$, sh -- subharmonic frequency without f$_{0.61}$, shc -- f$_{\rm{sh}}$+ f$_{\rm{1O}}$, p -- power excess near 0.5~f$_{0.61}$, af -- additional frequency, f$_{0.68}$-- additional signal at $P_{\rm{1O}}$/$P_{\rm{0.68}}$$\sim$0.68, mod? -- possible amplitude modulation) 
}
\\
    
      
TIC &  Sectors & $f_{\rm{1O}}$ & $A_{\rm{1O}}$ & $f_{\rm{0.61}}$ & $f_{\rm{sh}}$&  $f_{\rm{0.61}}$/ $f_{\rm{1O}}$&  $A_{\rm{0.61}}$/$A_{\rm{1O}}$&$A_{\rm{sh}}$/$A_{\rm{0.61}}$ & Remarks\\

\noalign{\vskip 2pt} 
\hline
\hline
\noalign{\vskip 2pt} 

4712990	&	S33+S34	&	0.34734(2)	&	0.0897(2)	&	0.5623(4)	&	0.2791(4)	&	0.6178(4)	&	0.037(1)	&	1.01(5)	&	c	\\
	&		&		&		&		&	0.2922(5)	&	--	&	--	&	--	&	sh	\\
4867143	&	S7	&	0.3097(1)	&	0.1008(6)	&	0.491(1)	&	0.249(1)	&	0.631(2)	&	0.049(3)	&	0.88(7)	&	c	\\
27344118	&	S19	&	0.36033(5)	&	0.0937(2)	&	--	&	0.2758(8)	&	--	&	--	&	--	&	sh	\\
	&	S59	&	0.36024(2)	&	0.09759(8)	&	0.5696(6)	&	0.2850(4)	&	0.6325(7)	&	0.0163(5)	&	1.69(6)	&	c	\\
28184016	&	S19	&	0.5994(2)	&	0.1000(7)	&	0.981(2)	&	0.544(3)	&	0.611(1)	&	0.035(3)	&	0.55(8)	&	c	\\
	&	S59	&	0.59978(5)	&	0.1119(2)	&	0.994(2)	&	0.5451(8)	&	0.604(1)	&	0.015(1)	&	2.3(2)	&	--	\\
28551922	&	S61+S62	&	0.341193(7)	&	0.09111(6)	&	0.5466(2)	&	0.2729(1)	&	0.6242(2)	&	0.0214(4)	&	1.99(4)	&	--	\\
	&		&		&		&	--	&	0.2516(3)	&	--	&	--	&	--	&	sh	\\
28907876	&	S19	&	0.5272(2)	&	0.1238(9)	&	0.840(2)	&	0.435(2)	&	0.628(1)	&	0.037(2)	&	1.1(1)	&	c	\\
	&		&		&		&	--	&	0.384(2)	&	--	&	--	&	--	&	sh	\\
	&	S59	&	0.52743(5)	&	0.1256(3)	&	0.839(2)	&	0.4008(5)	&	0.629(2)	&	0.011(1)	&	4.1(4)	&	--	\\
29455183	&	S43+S44+S45	&	0.47567(1)	&	0.0955(2)	&	0.78320(8)	&	0.3924(1)	&	0.60734(6)	&	0.0403(5)	&	0.50(1)	&	c, hif	\\
	&		&		&		&	0.7574(2)	&	--	&	0.6281(1)	&	0.0143(2)	&	--	&	--	\\
	&	S71+S72	&	0.47560(1)	&	0.0950(1)	&	0.7819(1)	&	--	&	0.6083(1)	&	0.0265(4)	&	--	&	c, p	\\
	&		&		&		&	0.7554(1)	&	--	&	0.6296(1)	&	0.0227(2)	&	--	&	--	\\
	&		&		&		&	0.7363(2)	&	--	&	0.6459(2)	&	0.0110(2)	&	--	&	--	\\
32837372	&	S6	&	0.3321(1)	&	0.0825(4)	&	0.543(1)	&	0.262(1)	&	0.611(1)	&	0.036(2)	&	1.19(9)	&	c	\\
	&	S33	&	0.33196(6)	&	0.0915(2)	&	0.5388(5)	&	0.2673(4)	&	0.6162(6)	&	0.063(2)	&	0.97(3)	&	c	\\
44753014	&	S43+S44+S45	&	0.385952(7)	&	0.09167(9)	&	0.6222(2)	&	0.3190(1)	&	0.6203(2)	&	0.0267(7)	&	1.76(5)	&	c	\\
	&		&		&		&	0.5998(2)	&	0.3009(2)	&	0.6435(2)	&	0.0188(7)	&	0.92(5)	&	c	\\
59617576	&	S6	&	0.31904(9)	&	0.0983(3)	&	0.505(2)	&	0.262(2)	&	0.632(3)	&	0.0081(7)	&	1.0(1)	&	f$_{0.68}$	\\
	&	S33	&	0.31864(4)	&	0.0979(2)	&	0.4967(5)	&	--	&	0.6416(7)	&	0.0212(5)	&	--	&	c, f$_{0.68}$		\\
	&	S43+S44+S55	&	0.318754(9)	&	0.1010(1)	&	0.5197(2)	&	0.2742(1)	&	0.6133(3)	&	0.0124(4)	&	2.17(8)	&	f$_{0.68}$		\\
	&	S71+S72	&	0.31884(1)	&	0.10049(9)	&	0.5266(2)	&	0.2662(1)	&	0.6055(2)	&	0.0228(3)	&	1.21(2)	&	c, f$_{0.68}$		\\
72285450	&	S19	&	0.3237(1)	&	0.1182(5)	&	0.529(1)	&	0.275(1)	&	0.612(1)	&	0.036(1)	&	0.51(4)	&	c	\\
	&	S59	&	0.32373(3)	&	0.1185(2)	&	0.5279(7)	&	0.2561(3)	&	0.6133(8)	&	0.0119(4)	&	3.3(1)	&	c	\\
81809914	&	S71+S72	&	0.46774(2)	&	0.1150(2)	&	0.7732(1)	&	0.3811(3)	&	0.6050(1)	&	0.0355(4)	&	0.41(1)	&	c	\\
83060637	&	S17+S18	&	0.27392(4)	&	0.0879(4)	&	0.4401(5)	&	--	&	0.6223(7)	&	0.041(2)	&	--	&	c, p	\\
	&		&		&		&	0.4573(8)	&	--	&	0.599(1)	&	0.018(1)	&	--	&	--	\\
	&	S24	&	0.2734(1)	&	0.0944(4)	&	0.453(1)	&	0.2007(6) 	&	0.604(2)	&	0.022(1)	&	2.4(2)	&	--	\\
	&	S57+S58	&	0.27384(1)	&	0.0915(1)	&	0.4377(2)	&	--	&	0.6257(3)	&	0.0264(5)	&	--	&	--	\\
	&		&		&		&	0.4613(2)	&	--	&	0.5936(2)	&	0.0209(5)	&	--	&	c	\\
	&		&		&		&	0.4232(2)	&	0.2052(1)	&	0.6471(2)	&	0.0185(4)	&	1.95(5)	&	--	\\
94867105	&	S7	&	0.38043(8)	&	0.1296(4)	&	0.618(1)	&	0.288(2)	&	0.616(1)	&	0.018(1)	&	0.78(8)	&	c	\\
	&	S34	&	0.38049(5)	&	0.1362(3)	&	0.616(1)	&	--	&	0.617(1)	&	0.0169(8)	&	--	&	c, p	\\
	&	S61	&	0.38010(3)	&	0.1348(2)	&	0.6123(5)	&	0.2904(5)	&	0.6208(5)	&	0.0246(5)	&	0.75(2)	&	c	\\
95586534	&	S7	&	0.41475(8)	&	0.0769(3)	&	0.6835(8)	&	--	&	0.6068(7)	&	0.0240(9)	&	--	&	c, mod?	\\
	&	S33	&	0.41478(5)	&	0.0835(2)	&	0.6798(5)	&	0.3499(9)	&	0.6102(5)	&	0.0231(6)	&	0.30(1)	&	c, mod?	\\
107403493	&	S7	&	0.5347(1)	&	0.0864(4)	&	0.859(1)	&	0.414(1)	&	0.6223(8)	&	0.054(3)	&	0.86(6)	&	c	\\
109208054	&	S34	&	0.52982(8)	&	0.0943(3)	&	0.834(1)	&	0.439(1)	&	0.6352(9)	&	0.045(2)	&	1.11(8)	&	c	\\
109722919	&	S7	&	0.4568(1)	&	0.1016(5)	&	0.761(2)	&	--	&	0.600(1)	&	0.014(1)	&	--	&	c, p	\\
	&	S34	&	0.45681(8)	&	0.1128(4)	&	0.755(1)	&	--	&	0.605(1)	&	0.026(1)	&	--	&	c	\\
	&	S61	&	0.45633(4)	&	0.1136(2)	&	0.7561(6)	&	--	&	0.6035(5)	&	0.0250(7)	&	--	&	c	\\
112414739	&	S7	&	0.3026(1)	&	0.0879(4)	&	0.483(1)	&	--	&	0.626(2)	&	0.028(1)	&	--	&	c	\\
	&		&		&		&	--	&	0.274(1) 	&	--	&	--	&	--	&	sh	\\
	&	S34	&	0.30257(7)	&	0.0919(3)	&	0.4898(6)	&	0.2518(4)	&	0.6177(7)	&	0.041(1)	&	1.80(5)	&	c	\\
	&	S61	&	0.30287(5)	&	0.0970(2)	&	0.4909(3)	&	0.2431(3)	&	0.6169(4)	&	0.0487(7)	&	1.44(3)	&	c	\\
113843324	&	S7	&	0.5723(1)	&	0.0920(5)	&	0.921(2)	&	0.483(1)	&	0.622(1)	&	0.036(3)	&	1.3(1)	&	--	\\
	&	S34	&	0.57214(9)	&	0.1021(4)	&	0.895(1)	&	--	&	0.6395(8)	&	0.042(2)	&	--	&	p	\\
	&	S61	&	0.57205(5)	&	0.1005(2)	&	--	&	0.488(1)	&	--	&	--	&	--	&	sh	\\
	&		&		&		&	--	&	0.4392(8)	&	--	&	--	&	--	&	sh	\\
114419152 &	S44+S45	&	0.35536(2)	&	0.1096(2)	&	--	&	0.2953(2)	&	--	&	--	&	--	&	shc	\\
&		&		&		&	--	&	0.2703(3)	&	--	&	--	&	--	&	sh	\\
    
	&	S71+S72	&	0.35527(1)	&	0.1150(1)	&	0.5716(2)	&	0.2949(2)	&	0.6215(4)	&	0.0134(4)	&	2.34(8)	&	c	\\
	&		&		&		&	--	&	0.2786(3)	&	--	&	--	&	--	&	sh	\\
116747560	&	S19	&	0.36601(9)	&	0.0866(3)	&	--	&	0.303(2)	&	--	&	--	&	--	&	sh	\\
	&	S44+S45	&	0.36598(2)	&	0.0857(1)	&	0.5964(4)	&	0.3257(3)	&	0.613(5)	&	0.0240(9)	&	1.5(1)	&	c, mod?	\\
	&		&		&		&	0.5737(4)	&	0.2948(4)	&	0.638(5)	&	0.0195(9)	&	1.2(1)	&	c	\\
	&	S59	&	0.36614(5)	&	0.0843(1)	&	0.5785(6)	&	0.3072(3)	&	0.633(6)	&	0.0360(9)	&	1.88(6)	&	c	\\
118592633	&	S43+S44	&	0.317556(8)	&	0.10981(8)	&	0.5013(3)	&	--	&	0.6335(4)	&	0.0130(4)	&	--	&	c	\\
	&	S70	&	0.31758(1)	&	0.11093(6)	&	0.5129(6)	&	--	&	0.6191(7)	&	0.0050(1)	&	--	&	c	\\
	&	S71	&	0.31765(2)	&	0.10767(7)	&	0.5127(4)	&	0.2534(4)	&	0.6196(5)	&	0.0153(3)	&	0.95(2)	&	c	\\
123734693	&	S6	&	0.5416(2)	&	0.0982(6)	&	0.906(1)	&	0.438(1)	&	0.598(1)	&	0.042(2)	&	0.96(7)	&	c	\\
	&		&		&		&	--	&	0.400(2)	&	--	&	--	&	--	&	sh	\\
	&	S33	&	0.54128(7)	&	0.1151(4)	&	0.887(1)	&	0.4439(6)	&	0.6101(8)	&	0.0105(5)	&	2.2(1)	&	c, hif	\\
	&		&		&		&	--	&	0.3955(7)	&	--	&	--	&	--	&	sh	\\
124702643	&	S7	&	0.4506(1)	&	0.0800(4)	&	0.752(3)	&	0.372(1)	&	0.599(2)	&	0.014(2)	&	3.3(5)	&	--	\\
	&		&		&		&	0.715(2)	&	--	&	0.630(2)	&	0.017(2)	&	--	&	--	\\
	&	S33	&	0.45096(6)	&	0.0996(3)	&	0.740(2)	&	0.3778(6)	&	0.609(1)	&	0.016(1)	&	2.8(2)	&	--	\\
126545806	&	S7	&	0.3056(1)	&	0.0854(4)	&	0.494(1)	&	0.2474(6)	&	0.619(1)	&	0.029(1)	&	2.1(1)	&	c, shc	\\
	&	S34	&	0.30532(7)	&	0.0844(3)	&	0.480(1)	&	0.2515(4)	&	0.636(2)	&	0.0146(9)	&	6.9(4)	&	shc	\\
	&	S61	&	0.30521(4)	&	0.0889(2)	&	0.4973(7)	&	0.2484(3)	&	0.6138(9)	&	0.0211(7)	&	3.9(1)	&	--	\\
131595172	&	S54	&	0.25294(9)	&	0.1187(5)	&	0.402(1)	&	0.209(1)	&	0.629(2)	&	0.0177(8)	&	1.06(7)	&	c	\\
	&	S81	&	0.25291(4)	&	0.1133(2)	&	0.3957(6)	&	--	&	0.6391(9)	&	0.0188(5)	&	--	&	c	\\
139446463	&	S7	&	0.4439(1)	&	0.0640(3)	&	0.720(2)	&	0.340(2)	&	0.617(1)	&	0.045(3)	&	0.69(9)	&	c	\\
145326073	&	S7	&	0.36782(5)	&	0.0812(2)	&	0.584(2)	&	0.302(1)	&	0.630(2)	&	0.011(1)	&	2.2(2)	&	--	\\
	&	S34	&	0.36773(3)	&	0.0985(1)	&	0.593(1)	&	0.319(1)	&	0.620(1)	&	0.0197(9)	&	0.60(5)	&	c	\\
	&	S61	&	0.36758(3)	&	0.0937(1)	&	0.5848(4)	&	0.3023(5)	&	0.6286(4)	&	0.060(1)	&	0.59(2)	&	c	\\
147184001	&	S6+S7	&	0.41388(4)	&	0.0938(3)	&	0.6688(4)	&	0.3410(4)	&	0.6189(4)	&	0.051(2)	&	1.21(6)	&	c, b	\\
	&	S33	&	0.41406(5)	&	0.1142(2)	&	0.6631(9)	&	0.3297(7)	&	0.6244(9)	&	0.031(1)	&	1.51(8)	&	c, b	\\
147372646	&	S6+S7	&	0.42994(3)	&	0.0887(3)	&	0.7221(5)	&	0.3730(5)	&	0.5954(4)	&	0.0206(9)	&	0.72(5)	&	c	\\
	&	S33	&	0.43025(5)	&	0.1054(2)	&	0.7145(6)	&	--	&	0.6021(5)	&	0.0246(6)	&	--	&	c	\\
148168449	&	S7	&	0.5533(2)	&	0.0941(7)	&	0.898(1)	&	0.433(2)	&	0.6161(8)	&	0.034(2)	&	0.51(5)	&	c	\\
	&	S33	&	0.55308(8)	&	0.1160(4)	&	0.893(1)	&	0.422(2)	&	0.6193(8)	&	0.021(1)	&	0.56(5)	&	c, hif	\\
148500691	&	S7	&	0.4728(1)	&	0.0972(5)	&	0.777(1)	&	--	&	0.6085(9)	&	0.028(1)	&	--	&	c	\\
	&		&		&		&	0.745(1)	&	0.3671(8)	&	0.634(1)	&	0.016(1)	&	2.6(2)	&	--	\\
	&	S33	&	0.47280(6)	&	0.1114(3)	&	0.756(1)	&	0.385(1)	&	0.625(1)	&	0.0110(7)	&	0.93(8)	&	c, hif	\\
	&		&	--	&		&	--	&	0.346(1)	&	--	&	--	&	--	&	--	\\
166922631	&	S33	&	0.53660(8)	&	0.1329(5)	&	0.894(1)	&	--	&	0.6002(8)	&	0.0133(7)	&	--	&	c, p	\\
	&	S71+S72	&	0.53650(2)	&	0.1312(2)	&	0.8849(4)	&	--	&	0.6063(2)	&	0.0156(5)	&	--	&	c, p	\\
	&		&		&		&	0.8609(5)	&	--	&	0.6232(4)	&	 0.0097(5)	&	--	&	--	\\
168597634	&	S7	&	0.36864(7)	&	0.1017(3)	&	0.610(1)	&	0.294(1)	&	0.604(1)	&	0.022(1)	&	0.69(5)	&	c	\\
	&	S33	&	0.36841(3)	&	0.1057(2)	&	0.5818(7)	&	0.2934(5)	&	0.6332(7)	&	0.0136(4)	&	1.65(6)	&	c	\\
202057552	&	S6	&	0.37331(9)	&	0.1070(4)	&	0.599(2)	&	0.288(1)	&	0.623(2)	&	0.034(2)	&	0.79(6)	&	c	\\
	&		&		&		&	--	&	0.336(2)	&	--	&	--	&	--	&	sh	\\
	&	S33	&	0.37303(3)	&	0.1160(2)	&	0.6112(7)	&	0.326(1)	&	0.6103(7)	&	0.0185(6)	&	0.66(4)	&	c	\\
	&		&		&		&	0.579(1)	&	--	&	0.643(1)	&	0.0120(6)	&	--	&	--	\\
206897061	&	S6	&	0.5124(2)	&	0.0909(6)	&	0.851(1)	&	--	&	0.602(1)	&	0.019(1)	&	--	&	c	\\
	&	S33	&	0.51246(7)	&	0.1068(4)	&	0.8491(9)	&	--	&	0.6035(6)	&	0.0117(5)	&	--	&	c	\\
229163413	&	S33	&	0.32516(6)	&	0.0901(2)	&	0.5183(5)	&	0.261(2)	&	0.6274(6)	&	0.049(1)	&	0.26(2)	&	c	\\
232335529	&	S33	&	0.38634(5)	&	0.1409(3)	&	0.6201(8)	&	0.3031(5)	&	0.6230(8)	&	0.0241(9)	&	1.92(8)	&	c	\\
234218727	&	S6	&	0.4324(1)	&	0.1192(6)	&	0.712(2)	&	--	&	0.608(1)	&	0.017(1)	&	--	&	c	\\
	&	S33	&	0.43254(6)	&	0.1235(3)	&	0.7186(5)	&	--	&	0.6019(4)	&	0.0243(5)	&	--	&	c	\\
234943347	&	S6	&	0.4761(1)	&	0.0696(3)	&	0.763(2)	&	0.375(1)	&	0.624(2)	&	0.027(2)	&	1.6(1)	&	c, hif	\\
251182052	&	S6	&	0.3288(1)	&	0.0939(5)	&	--	&	0.277(2)	&	--	&	--	&	--	&	--	\\
	&	S33	&	0.32906(7)	&	0.1305(4)	&	0.5342(6)	&	0.2743(9)	&	0.6160(7)	&	0.067(2)	&	0.37(2)	&	c, f$_{0.68}$		\\
251252233	&	S18	&	0.37785(7)	&	0.0810(2)	&	0.617(2)	&	0.333(1)	&	0.612(2)	&	0.017(1)	&	1.7(1)	&	c	\\
	&	S58	&	0.37814(2)	&	0.08018(6)	&	0.6305(2)	&	0.3331(5)	&	0.5998(2)	&	0.0267(2)	&	0.49(1)	&	c	\\
251631993	&	S58	&	0.32715(3)	&	0.1048(1)	&	0.5251(6)	&	0.2794(3)	&	0.6230(7)	&	0.0147(5)	&	2.21(8)	&	c	\\
&		&		&		&	--	&	0.2375(5)	&	--	&	--	&	--	&	sh	\\
255977934	&	S18	&	0.40364(8)	&	0.0862(3)	&	0.646(2)	&	0.326(1)	&	0.625(2)	&	0.021(1)	&	1.5(1)	&	f$_{0.68}$		\\
	&	S24+S25	&	0.40345(2)	&	0.0850(2)	&	0.6485(8)	&	0.3349(3)	&	0.6221(7)	&	0.014(1)	&	3.2(3)	&	f$_{0.68}$	\\
	&		&		&		&	--	&	0.2971(6)	&	--	&	--	&	--	&	sh	\\
	&		&		&		&	--	&	0.3564(6)	&	--	&	--	&	--	&	sh	\\
	&	S58	&	0.40343(1)	&	0.08538(6)	&	0.6530(6)	&	0.3294(8)	&	0.6178(6)	&	0.0135(4)	&	0.70(3)	&	c, f$_{0.68}$, af	\\
258935093	&	S12	&	0.32624(5)	&	0.0862(2)	&	0.5351(7)	&	0.266(1)	&	0.6097(8)	&	0.0240(8)	&	0.36(3)	&	c	\\
	&	S38+S39	&	0.32625(1)	&	0.0958(1)	&	0.5455(3)	&	--	&	0.598(3)	&	0.0188(5)	&	--	&	c	\\
	&		&		&		&	0.5033(5)	&	--	&	0.6482(6)	&	0.0088(4)	&	--	&	--	\\
	&	S65	&	0.32616(1)	&	0.09524(7)	&	0.5421(4)	&	0.2625(5)	&	0.6017(5)	&	0.0131(3)	&	0.66(2)	&	c	\\
260292560	&	S16+S17	&	0.32556(5)	&	0.0853(3)	&	0.5340(9)	&	0.2652(5)	&	0.610(1)	&	0.031(2)	&	1.8(2)	&	c	\\
	&	S57	&	0.32548(3)	&	0.0884(1)	&	0.531(1)	&	0.2770(5)	&	0.612(1)	&	0.019(1)	&	2.5(2)	&	c	\\
	&		&		&		&	--	&	0.2456(4)	&	--	&	--	&	--	&	sh	\\
	&	S76+S77	&	0.32555(1)	&	0.0913(1)	&	0.5299(3)	&	0.2765(2)	&	0.6144(3)	&	0.040(1)	&	1.26(5)	&	c	\\
	&		&		&		&	--	&	0.2475(3)	&	--	&	--	&	--	&	sh	\\
270729245	&	S17+S18	&	0.40202(3)	&	0.0684(2)	&	0.6658(5)	&	0.3546(9)	&	0.6038(5)	&	0.027(1)	&	0.53(5)	&	c, b	\\
	&	S24	&	0.40206(8)	&	0.0674(2)	&	0.670(1)	&	0.348(1)	&	0.600(1)	&	0.029(1)	&	0.52(5)	&	c, b	\\
	&	S57+S58	&	0.40203(1)	&	0.07247(7)	&	0.6713(2)
	&	0.3450(3)	&	0.599(2)	&	0.0345(7)	&	0.49(2)	&	c, b	\\
272999674	&	S19	&	0.4968(1)	&	0.0832(4)	&	0.793(1)	&	0.4044(8)	&	0.6262(8)	&	0.057(2)	&	1.56(9)	&	c, hif	\\
	&	S59	&	0.49642(4)	&	0.0824(1)	&	0.791(1)	&	0.4007(3)	&	0.6273(8)	&	0.028(1)	&	3.1(1)	&	c, hif, shc	\\
280536157	&	S6	&	0.3602(1)	&	0.1317(5)	&	0.565(2)	&	0.272(3)	&	0.638(2)	&	0.030(2)	&	0.50(7)	&	c	\\
	&	S33	&	0.36024(4)	&	0.1322(2)	&	0.559(1)	&	0.2801(6)	&	0.645(1)	&	0.0137(7)	&	2.4(1)	&	c	\\
283374564	&	S12	&	0.2285(1)	&	0.1074(7)	&	0.3579(9)	&	0.171(1)	&	0.638(2)	&	0.042(2)	&	0.50(4)	&	c	\\
	&	S38+S39	&	0.22854(3)	&	0.1072(3)	&	0.3587(2)	&	0.1804(3)	&	0.6371(3)	&	0.0437(7)	&	0.43(2)	&	c	\\
	&	S65	&	0.22835(4)	&	0.1062(2)	&	0.3619(3)	&	--	&	0.6309(5)	&	0.0253(3)	&	--	&	c, p	\\
286048485	&	S44+S45	&	0.81796(4)	&	0.1658(7)	&	1.3373(9)	&	0.6690(9)	&	0.6116(4)	&	0.0096(8)	&	1.1(1)	&	hif, mod?	\\
	&	S71	&	0.81794(8)	&	0.1659(5)	&	1.3343(9)	&	--	&	0.6130(4)	&	0.0243(9)	&	--	&	c	\\
288899618	&	S7	&	0.38482(8)	&	0.1325(5)	&	0.621(2)	&	0.309(1)	&	0.620(2)	&	0.017(1)	&	1.2(1)	&	c	\\
	&	S34	&	0.38460(5)	&	0.1388(3)	&	0.610(2)	&	0.308(1)	&	0.630(2)	&	0.0088(6)	&	1.3(1)	&	c	\\
306533156	&	S7	&	0.5064(1)	&	0.0797(3)	&	0.805(2)	&	0.4270(9)	&	0.629(1)	&	0.025(2)	&	1.9(2)	&	c	\\
	&		&		&		&	--	&	0.390(2)	&	--	&	--	&	--	&	sh	\\
	&	S33	&	0.50665(5)	&	0.0986(2)	&	0.809(2)	&	0.425(2)	&	0.626(2)	&	0.010(1)	&	1.4(2)	&	c	\\
	&		&		&		&	--	&	0.368(1)	&	--	&	--	&	--	&	sh	\\
306651503	&	S7	&	0.2737(1)	&	0.0963(6)	&	0.426(2)	&	0.218(1)	&	0.643(2)	&	0.024(2)	&	1.5(1)	&	c	\\
	&	S33+S34	&	0.27377(3)	&	0.1029(3)	&	0.4324(2)	&	0.2280(5)	&	0.6332(4)	&	0.045(1)	&	0.47(2)	&	c	\\
	&		&		&		&	--	&	0.2070(6)	&	--	&	--	&	--	&	sh	\\
312060948	&	S17+S18	&	0.35408(5)	&	0.1265(6)	&	0.5569(7)	&	0.2768(6)	&	0.6358(8)	&	0.031(2)	&	1.01(8)	&	c	\\
	&	S58	&	0.35393(4)	&	0.1198(2)	&	0.5554(6)	&	0.2894(6)	&	0.6372(7)	&	0.0271(8)	&	1.07(4)	&	c	\\
316965468	&	S18+S19	&	0.27159(4)	&	0.0751(2)	&	0.4279(4)	&	0.2232(4)	&	0.6348(7)	&	0.026(1)	&	0.86(5)	&	c, af	\\
	&	S59	&	0.27167(3)	&	0.0764(1)	&	0.4253(2)	&	--	&	0.6388(3)	&	0.0422(4)	&	--	&	c, af	\\
317210786	&	S11	&	0.2353(2)	&	0.0799(7)	&	0.3682(8)	&	0.187(2)	&	0.639(1)	&	0.069(2)	&	0.19(2)	&	c	\\
	&	S65	&	0.23517(6)	&	0.0902(2)	&	0.3678(4)	&	--	&	0.6394(7)	&	0.0472(8)	&	--	&	c	\\
331985937	&	S18	&	0.47612(8)	&	0.0758(3)	&	0.773(2)	&	0.381(2)	&	0.616(1)	&	0.036(3)	&	1.1(1)	&	c, hif	\\
	&	S58	&	0.47573(3)	&	0.0857(1)	&	0.7626(4)	&	0.3752(4)	&	0.6238(3)	&	0.069(1)	&	1.02(3)	&	c	\\
333611596	&	S6	&	0.3066(1)	&	0.0907(4)	&	0.503(2)	&	0.251(1)	&	0.609(2)	&	0.021(1)	&	1.9(1)	&	c	\\
	&	S33	&	0.30630(6)	&	0.1002(3)	&	0.4970(5)	&	0.2306(5)	&	0.6163(6)	&	0.067(2)	&	0.77(3)	&	c	\\
	&		&		&		&	--	&	0.2716(7)	&	--	&	--	&	--	&	sh	\\
337991027	&	S16	&	0.32755(5)	&	0.0964(2)	&	0.516(1)	&	0.278(1)	&	0.635(2)	&	0.0126(7)	&	1.02(8)	&	c	\\
	&	S17	&	0.32769(5)	&	0.1030(2)	&	--	&	0.2590(7)	&	--	&	--	&	--	&	sh	\\
	&	S24	&	0.32758(4)	&	0.1036(2)	&	--	&	0.2519(7)	&	--	&	--	&	--	&	sh	\\
	&	S57	&	0.32765(1)	&	0.10447(7)	&	--	&	0.2466(2)	&	--	&	--	&	--	&	sh	\\
338152926	&	S64	&	0.49301(4)	&	0.0952(2)	&	0.8016(3)	&	0.3981(5)	&	0.6150(3)	&	0.064(1)	&	0.61(2)	&	c, b	\\
341556329	&	S16	&	0.35884(9)	&	0.0980(4)	&	0.578(4)	&	0.261(1)	&	0.620(4)	&	0.012(2)	&	3.7(7)	&	c, af	\\
	&	S56	&	0.35912(3)	&	0.1089(2)	&	0.564(1)	&	0.3057(9)	&	0.636(2)	&	0.016(1)	&	1.6(1)	&	c, af	\\
	&	S76	&	0.35899(3)	&	0.1070(2)	&	0.5740(9)	&	0.3019(8)	&	0.625(1)	&	0.031(1)	&	1.26(7)	&	c	\\
358427670	&	S16	&	0.2586(1)	&	0.0813(4)	&	--	&	0.1920(7)	&	--	&	--	&	--	&	sh	\\
	&	S55+S56+S67	&	0.258911(6)	&	0.0847(1)	&	0.4148(1)	&	0.20639(6)	&	0.6242(2)	&	0.0053(1)	&	2.60(6)	&	c	\\
	&	S76	&	0.25918(3)	&	0.0841(1)	&	--	&	0.2135(2)	&	--	&	--	&		&	sh	\\
365119871	&	S16	&	0.35667(8)	&	0.0796(3)	&	0.584(1)	&	0.288(1)	&	0.610(1)	&	0.034(2)	&	0.72(6)	&	c	\\
	&	S56	&	0.35692(3)	&	0.0860(1)	&	0.5912(6)	&	0.2878(5)	&	0.6037(6)	&	0.0300(9)	&	1.36(5)	&	c	\\
    &		&		&		&	--	&	0.3216(6)	&	--	&	--	&	--		sh	\\
	&	S75+S76	&	0.35688(1)	&	0.08471(9)	&	0.5883(5)	&	0.3044(5)	&	0.6066(5)	&	0.0162(8)	&	0.98(7)	&	c	\\
    &		&		&		&	--	&	0.3261(2)	&	--	&	--	&	--	&	sh	\\
365297180	&	S15+S16	&	0.31214(3)	&	0.0991(3)	&	--	&	0.2466(4)	&	--	&	--	&	--	&	sh, f$_{0.68}$		\\
	&	S56+S57	&	0.312177(7)	&	0.09536(7)	&	0.5011(1)	&	0.2497(2)	&	0.6230(2)	&	0.0335(5)	&	0.58(2)	&	c, f$_{0.68}$		\\
	&		&		&		&	0.4774(2)	&	--	&	0.6539(2)	&	0.0225(5)	&	--	&	c	\\
	&	S76+S77	&	0.312075(8)	&	0.09989(8)	&	0.4952(3)	&	0.2419(2)	&	0.6301(3)	&	0.0216(6)	&	1.06(4)	&	c, f$_{0.68}$		\\
	&		&		&		&	0.4787(2)	&	--	&	0.6520(4)	&	0.0167(5)	&	--	&	--	\\
365535966	&	S56+S57	&	0.51227(2)	&	0.0931(2)	&	0.8585(9)	&	0.4176(7)	&	0.5967(5)	&	0.015(1)	&	1.3(1)	&	c	\\
386157925	&	S9+S10	&	0.24306(4)	&	0.0820(3)	&	0.3889(6)	&	0.1948(4)	&	0.6249(9)	&	0.021(1)	&	1.8(1)	&	c	\\
	&	S35+S36	&	0.24278(3)	&	0.0913(2)	&	--	&	0.1921(3)	&	---	&	--	&	--	&	--	\\
	&	S62+S63	&	0.24295(1)	&	0.0909(1)	&	0.3849(2)	&	0.1980(2)	&	0.6312(2)	&	0.0209(2)	&	1.03(2)	&	c	\\
&		&		&		&	--	&	0.1737(2)	&	--	&	--	&	--	&	sh	\\
395471265	&	S54	&	0.24922(5)	&	0.0719(2)	&	0.3952(7)	&	--	&	0.631(1)	&	0.0183(6)	&	--	&	c, f$_{0.68}$		\\
	&	S80+S81	&	0.24905(1)	&	0.07280(8)	&	0.3906(1)	&	0.1956(2)	&	0.6377(2)	&	0.0440(6)	&	0.48(1)	&	c, f$_{0.68}$		\\
399269858	&	S18	&	0.4732(1)	&	0.1167(7)	&	0.7744(5)	&	--	&	0.6110(4)	&	0.055(1)	&	--	&	c	\\
	&	S24	&	0.4732(1)	&	0.1175(6)	&	0.760(1)	&	0.391(1)	&	0.623(1)	&	0.0127(7)	&	0.69(6)	&	c	\\
	&	S25	&	0.4731(1)	&	0.1194(8)	&	0.7748(8)	&	--	&	0.6106(7)	&	0.027(1)	&	--	&	c	\\
	&	S52	&	0.47320(8)	&	0.1199(4)	&	0.7835(6)	&	--	&	0.6040(5)	&	0.0248(6)	&	--	&	c	\\
	&	S58	&	0.47327(3)	&	0.1169(2)	&	0.7816(2)	&	--	&	0.6055(2)	&	0.0275(3)	&	--	&	c	\\
399437508	&	S58	&	0.31011(2)	&	0.1133(1)	&	0.4864(3)	&	0.2317(2)	&	0.6376(4)	&	0.0299(4)	&	0.78(2)	&	c	\\
400163703	&	S19	&	0.34521(9)	&	0.0944(4)	&	0.554(2)	&	0.281(1)	&	0.623(2)	&	0.022(2)	&	1.6(2)	&	c	\\
&		&		&		&	--	&	0.2473(9)	&	--	&	--	&	--	&	sh	\\
	&	S43+S44+S45	&	0.344754(6)	&	0.09707(9)	&	0.5521(2)	&	0.2713(2)	&	0.6245(2)	&	0.0150(4)	&	1.46(5)	&	c	\\
    &		&		&		&	--	&	0.2522(1)	&	--	&	--	&	--	&	sh	\\
	&	S72	&	0.34470(2)	&	0.10100(9)	&	0.5296(5)	&	0.2673(4)	&	0.6509(5)	&	0.0193(4)	&	1.29(4)	&	c	\\
	&		&		&		&	0.5720(7)	&	--	&	0.6026(7)	&	0.0128(4)	&	--	&	c	\\
407021836	&	S17+S18	&	0.38702(3)	&	0.0912(3)	&	0.605(1)	&	0.3073(9)	&	0.639(1)	&	0.022(2)	&	1.1(1)	&	c	\\
	&	S24	&	0.38739(8)	&	0.0949(4)	&	0.633(2)	&	0.3109(7)	&	0.612(2)	&	0.021(2)	&	3.5(4)	&	c	\\
	&	S58	&	0.38700(2)	&	0.0883(1)	&	0.6187(7)	&	0.325(1)	&	0.6255(7)	&	0.028(1)	&	0.65(4)	&	c	\\
411946994	&	S38	&	0.23486(9)	&	0.0880(4)	&	0.3786(8)	&	0.1882(5)	&	0.620(1)	&	0.034(1)	&	1.75(8)	&	c, af	\\
	&	S64	&	0.23480(5)	&	0.0940(2)	&	0.3762(6)	&	0.1915(7)	&	0.624(1)	&	0.038(1)	&	0.79(3)	&	c	\\
412638643	&	S16+S17	&	0.47304(5)	&	0.1085(5)	&	0.7524(4)	&	0.3760(3)	&	0.6287(4)	&	0.0236(9)	&	1.55(7)	&	c	\\
	&	S24	&	0.4730(1)	&	0.1082(6)	&	0.7546(8)	&	0.3768(5)	&	0.6268(7)	&	0.0221(9)	&	1.88(9)	&	c, hif	\\
	&	S56+S57	&	0.47304(1)	&	0.1101(1)	&	0.75110(9)	&	0.37556(7)	&	0.62980(8)	&	0.0229(2)	&	1.59(2)	&	c, hif	\\
	&	S76+S77	&	0.47294(1)	&	0.1089(2)	&	0.75088(7)	&	0.37458(9)	&	0.62984(6)	&	0.0397(3)	&	1.11(1)	&	c	\\
421389234	&	S58	&	0.42801(3)	&	0.1079(2)	&	0.684(1)	&	0.3410(7)	&	0.626(1)	&	0.019(1)	&	1.5(1)	&	c	\\
424326261	&	S41	&	0.31614(3)	&	0.0882(1)	&	0.5248(7)	&	--	&	0.602(8)	&	0.0277(9)	&	--	&	c	\\
	&	S54+S55	&	0.31594(2)	&	0.0809(1)	&	0.4927(6)	&	0.2479(4)	&	0.6413(8)	&	0.016(1)	&	1.7(1)	&	--	\\
	&	S81	&	0.31613(2)	&	0.08806(9)	&	0.5329(7)	&	0.2521(8)	&	0.5933(8)	&	0.0268(9)	&	0.81(4)	&	c	\\
433877426	&	S24	&	0.33006(4)	&	0.1017(2)	&	0.536(1)	&	0.2768(8)	&	0.616(1)	&	0.0093(5)	&	1.7(1)	&	c	\\
	&	S57	&	0.33021(2)	&	0.10606(9)	&	0.5577(2)	&	0.3064(2)	&	0.5921(2)	&	0.0200(2)	&	0.91(2)	&	c	\\
436534765	&	S33	&	0.67753(7)	&	0.0800(3)	&	1.1156(6)	&	0.6292(5)	&	0.6073(3)	&	0.0192(5)	&	1.29(5)	&	c	\\
440962293	&	S9	&	0.4432(2)	&	0.0604(4)	&	0.694(3)	&	0.363(3)	&	0.639(3)	&	0.040(5)	&	1.3(2)	&	c	\\
&		&		&		&	--	&	0.317(3)	&	--	&	--	&	--	&	sh	\\
	&	S37	&	0.4432(1)	&	0.0714(3)	&	0.715(1)	&	0.357(1)	&	0.620(1)	&	0.067(4)	&	0.88(7)	&	c	\\
	&	S62+S63	&	0.44300(2)	&	0.0893(2)	&	0.7075(4)	&	0.3574(5)	&	0.6261(3)	&	0.060(2)	&	0.74(4)	&	c	\\
450463355	&	S11	&	0.4343(1)	&	0.0794(4)	&	0.7154(6)	&	0.370(2)	&	0.6071(5)	&	0.041(1)	&	0.29(2)	&	c	\\
	&	S37+S38	&	0.43371(2)	&	0.0858(1)	&	0.6677(3)	&	0.3327(3)	&	0.6495(3)	&	0.0161(4)	&	0.88(3)	&	c	\\
	&		&		&		&	0.7180(2)	&	0.3563(2)	&	0.6041(2)	&	0.0115(4)	&	0.92(4)	&	--	\\
	&	S64+S65	&	0.433765(8)	&	0.08466(7)	&	0.6657(2)	&	0.3348(2)	&	0.6516(2)	&	0.0096(2)	&	1.49(3)	&	--	\\
	&		&		&		&	0.7167(1)	&	0.3595(2)	&	0.6053(1)	&	0.0134(2)	&	0.73(2)	&	c	\\
463918793	&	S9+S10	&	0.35204(2)	&	0.0892(2)	&	0.5893(5)	&	--	&	0.5974(5)	&	0.026(1)	&	--	&	c	\\
	&	S36+S37	&	0.35207(1)	&	0.0985(1)	&	0.5941(2)	&	--	&	0.5926(2)	&	0.0249(5)	&	--	&	--	\\
	&		&		&		&	0.5718(5)	&	0.2682(3)	&	0.6157(6)	&	0.0085(4)	&	2.6(1)	&	c	\\
	&	S63+S64	&	0.352103(7)	&	0.10113(6)	&	0.5878(1)	&	0.2875(2)	&	0.5990(1)	&	0.0308(2)	&	0.269(8)	&	c	\\
	&		&		&		&	0.5431(2)	&	--	&	0.6483(2)	&	0.0093(2)	&	--	&	--	\\
722289976	&	S44	&	0.58120(9)	&	0.1167(4)	&	0.906(2)	&	0.451(2)	&	0.641(2)	&	0.015(1)	&	1.1(1)	&	f$_{0.68}$		\\
	&	S45	&	0.58035(8)	&	0.1211(4)	&	0.943(2)	&	0.475(1)	&	0.615(1)	&	0.019(2)	&	1.8(2)	&	c f$_{0.68}$		\\
&		&		&		&	--	&	0.538(2)	&	--	&	--	&	--	&	sh	\\

     \hline
\label{tab:061}

\end{longtable}

\clearpage

\begin{table*}
    \caption{Stars with other additional frequencies.   (c -- combination frequency, af -- additional frequency, 0.61 -- additional signal in the $P_{\rm{0.61}}$/$P_{\rm{1O}}$: 0.60--0.65 range.) }
    \centering
    \begin{tabular}{lllllllll}

TIC &  Sectors & $f_{\rm{1O}}$ & $A_{\rm{1O}}$ & $f_{\rm{x}}$  &  $A_{\rm{x}}$ & $f_{\rm{x}}$/ $f_{\rm{1O}}$&  $A_{\rm{x}}$/$A_{\rm{1O}}$& Remarks \\

\noalign{\vskip 2pt} 
\hline
\hline
\noalign{\vskip 2pt} 

32664565	&	S6	&	0.8697(2)	&	0.0696(5)	&	0.7444(8)	&	0.0030(1)	&	0.8559(9)	&	0.043(1)	&	c, sh?	\\
	&	S33	&	0.86976(8)	&	0.0759(3)	&	0.7439(6)	&	0.00309(9)	&	0.8553(7)	&	0.041(1)	&	c, sh?	\\
50032287	&	S7	&	0.2948(1)	&	0.0941(5)	&	0.249(1)	&	0.0024(1)	&	0.845(3)	&	0.026(1)	&	p, sh?	\\
	&	S61	&	0.29472(4)	&	0.1034(2)	&	0.2587(4)	&	0.00529(8)	&	0.878(1)	&	0.0512(8)	&	p, sh?	\\
54855309	&	S43+S44+S45	&	0.66123(2)	&	0.0997(2)	&	0.60543(8)	&	0.00347(4)	&	0.9156(1)	&	0.0348(4)	&	--	\\
	&	S71+S71	&	0.66120(2)	&	0.0947(1)	&	0.6061(1)	&	0.00362(3)	&	0.9167(2)	&	0.0382(3)	&	--	\\
142523103	&	S7	&	0.41736(5)	&	0.0588(1)	&	0.5915(6)	&	0.00231(6)	&	1.417(1)	&	0.039(1)	&	--	\\
	&	S33	&	0.41666(3)	&	0.06438(9)	&	0.5950(4)	&	0.00247(5)	&	1.4280(10)	&	0.0384(8)	&	--	\\
148871409	&	S61	&	0.34622(6)	&	0.1098(3)	&	0.2645(3)	&	0.00836(9)	&	0.7640(9)	&	0.0761(8)	&	c, sh?	\\
182161202	&	S7+S8	&	0.21343(6)	&	0.0799(4)	&	0.1612(3)	&	0.0048(1)	&	0.755(1)	&	0.060(1)	&	p, sh?	\\
	&	S34+S35	&	0.21297(4)	&	0.0930(3)	&	0.1630(1)	&	0.00557(7)	&	0.7654(5)	&	0.0599(8)	&	sh?	\\
	&	S61+S62	&	0.21275(2)	&	0.0865(2)	&	0.1613(1)	&	0.00917(8)	&	0.7582(5)	&	0.106(1)	&	sh?	\\
239378995	&	S55+S56	&	0.22368(2)	&	0.0593(1)	&	0.4856(2)	&	0.00176(4)	&	2.1710(9)	&	0.0297(7)	&	f$_{0.68}$	\\
	&	S75+S76	&	0.22366(1)	&	0.05964(9)	&	0.5127(2)	&	0.00182(4)	&	2.2923(9)	&	0.0305(7)	&	f$_{0.68}$	\\
255977934	&	S58	&	0.40343(1)	&	0.08538(6)	&	0.3600(5)	&	 0.00140(4)	&	 0.892(1)	&	 0.0164(5)	&	f$_{0.61}$, f$_{0.68}$, 2$f_{\rm{x}}$ 	\\
260101277	&	S19	&	0.7686(1)	&	0.0445(2)	&	0.6615(5)	&	0.00352(8)	&	0.8607(7)	&	0.079(2)	&	not in S59	\\
281312636	&	S6	&	0.6011(2)	&	0.0979(8)	&	0.490(1)	&	0.0040(2)	&	0.815(2)	&	0.041(2)	&	--	\\
	&		&		&		&	2.039(3)	&	0.0018(2)	&	3.392(5)	&	0.018(2)	&	c	\\
301602622	&	S6	&	1.2262(2)	&	0.0772(6)	&	0.791(1)	&	0.0020(1)	&	0.6451(8)	&	0.026(1)	&	--	\\
	&	S33	&	1.22638(8)	&	0.0860(3)	&	0.7919(4)	&	0.00229(5)	&	0.6457(3)	&	0.0266(6)	&	c	\\
310874217	&	S37+s38	&	0.20070(3)	&	0.0680(2)	&	0.1485(3)	&	0.0036(1)	&	0.740(1)	&	0.053(1)	&	sh?	\\
	&	S64+s65	&	0.20076(2)	&	0.0694(1)	&	0.1519(3)	&	0.00316(8)	&	0.757(1)	&	0.046(1)	&	sh?	\\
316965468	&	S18+S19	&	0.27159(4)	&	0.0751(2)	&	0.1706(4)	&	0.00236(9)	&	 0.628(1)	&	 0.031(1)	&	f$_{0.61}$	\\
	&	S59	&	0.27167(3)	&	0.0764(1)	&	0.1793(2)	&	 0.00417(4)	&	 0.6600(7)	&	 0.0546(5)	&	f$_{0.61}$	\\
337084951	&	S43+S44+S45	&	0.20667(2)	&	0.0746(2)	&	0.16763(4)	&	0.00724(4)	&	0.8111(2)	&	0.0971(6)	&	c, sh?	\\
	&	S71+S72	&	0.20654(2)	&	0.0714(1)	&	0.16779(3)	&	0.00813(2)	&	0.8124(2)	&	0.1139(3)	&	c, sh?	\\
341556329	&	S16	&	0.35884(9)	&	0.0980(4)	&	0.203(3)	&	 0.0017(2)	&	 0.566(8)	&	 0.017(2)	&	f$_{0.61}$	\\
	&	S56	&	0.35912(3)	&	0.1089(2)	&	0.182(1)	&	 0.0022(1)	&	 0.507(3)	&	 0.0202(9)	&	f$_{0.61}$	\\
356616458	&	S55+S56	&	0.17571(2)	&	0.0642(1)	&	0.1378(1)	&	0.00214(2)	&	0.7842(6)	&	0.0333(3)	&	sh?	\\
	&	S75+S76	&	0.17596(2)	&	0.0633(1)	&	0.13794(8)	&	0.00198(2)	&	0.7839(5)	&	0.0313(3)	&	sh?	\\
387405861	&	S16+S17+S18	&	0.25468(2)	&	0.1067(3)	&	0.2038(3)	&	0.00200(9)	&	0.800(1)	&	0.0187(8)	&	sh?	\\
	&	S24+S25	&	0.25463(4)	&	0.1086(4)	&	0.1835(3)	&	0.00242(7)	&	0.721(1)	&	0.0223(6)	&	sh?	\\
	&		&		&		&	0.2093(4)	&	0.00165(6)	&	0.822(2)	&	0.0152(6)	&	sh?	\\
	&	S52	&	0.25459(7)	&	0.1091(3)	&	0.1900(5)	&	0.00285(6)	&	0.746(2)	&	0.0261(6)	&	sh?	\\
	&	S56+57	&	0.25462(1)	&	0.1113(1)	&	0.1839(2)	&	0.00182(3)	&	0.7223(8)	&	0.0164(3)	&	sh?	\\
391416079	&	S19	&	0.3254(1)	&	0.0893(5)	&	0.1617(5)	&	0.0078(2)	&	0.497(2)	&	0.087(2)	&	--	\\
	&	S59	&	0.32597(4)	&	0.0845(1)	&	0.1620(2)	&	0.00756(6)	&	0.4970(6)	&	0.0895(7)	&	--	\\
407700748	&	S6	&	0.4551(1)	&	0.0477(2)	&	0.382(1)	&	0.00190(7)	&	0.839(2)	&	0.040(1)	&	sh?	\\
	&	&		&		&	0.328(2)	&	0.00077(7)	&	0.721(4)	&	0.016(1)	&	sh?	\\
	&	S33	&	0.45533(5)	&	0.0532(1)	&	0.3298(8)	&	0.00168(6)	&	0.724(2)	&	0.032(1)	&	sh?	\\
	&		&		&		&	0.400(1)	&	0.00118(6)	&	0.878(2)	&	0.022(1)	&	sh?	\\
411946994	&	S38	&	0.23486(9)	&	0.0880(4)	&	0.1550(8)	&	 0.0034(1)	&	 0.660(3)	&	 0.039(1)	&	f$_{0.61}$	\\
451049858	&	S37+S38	&	0.19896(3)	&	0.0717(2)	&	0.15149(7)	&	0.00929(6)	&	0.7614(4)	&	0.1296(9)	&	c, sh?\\
	&	S64+S65	&	0.19915(2)	&	0.0750(1)	&	0.15278(6)	&	0.00641(3)	&	0.7672(3)	&	0.0855(4)	&	c, sh?	\\
468619216	&	S55+S56	&	0.26452(1)	&	0.0803(1)	&	0.2261(2)	&	0.00251(6)	&	0.8548(8)	&	0.0313(7)	&	sh?	\\
	&	S75+S76	&	0.26430(1)	&	0.08123(9)	&	0.2270(3)	&	0.00221(6)	&	0.859(1)	&	0.0272(7)	&	p, sh?	\\

     \hline

    \end{tabular}
         
    \label{tab:add}
   
\end{table*}

\end{document}